\documentclass[preprint,prd,amsmath,amssymb,aps]{revtex4-1}

\usepackage{graphicx}
\usepackage{color}
\usepackage{rotating}
\usepackage{bm}
\usepackage{natbib}
\def\nat{Nature\ }
\def\aap{Astron.\ Astrophys.\ }
\def\apj{Astrophys.\ J.\ }
\def\apjl{Astrophys.\ J.\ Lett.\ }
\def\apjs{Astrophys.\ J.\ Supp.\ }
\def\aj{Astron.\ J.\ }
\def\mnras{Mon.\ Not.\ Roy.\ Astron.\ Soc.\ }

\def\prd{Phys.\ Rev.\ D\ }

\def\apss{Astrophys.\ Space\ Sci.\ }
\def\araa{Annu.\ Rev.\ Astron.\ Astrophys.\ }
\def\jcap{J.\ Cosmol.\ Astropart.\ Phys.\ }
\def\ssr{Space\ Sci.\ Rev.\ }
\def\pasj{Publications of the Astronomical Society of Japan}
\def\pasp{Publications of the Astronomical Society of Pacific}

\begin{document}
\title{Perspective on the cosmic-ray electron spectrum above TeV}

\author{Kun Fang}
\author{Bing-Bing Wang}
\author{Xiao-Jun Bi}
\author{Su-Jie Lin}
\author{Peng-Fei Yin}

\affiliation{Key Laboratory of Particle Astrophysics,
Institute of High Energy Physics, Chinese Academy of Sciences,
Beijing 100049, China}

\date{\today}

\begin{abstract}
The AMS-02 has measured the cosmic ray electron (plus positron) spectrum
up to $\sim$TeV with an unprecedent precision. The spectrum can be well described by
a power law without any obvious features above 10 GeV. The satellite instrument Dark
Matter Particle Explorer (DAMPE), which was launched a year ago, will measure
the electron spectrum up to 10 TeV with a high energy resolution.
The cosmic electrons beyond TeV may be attributed to few local
cosmic ray sources, such as supernova remnants. Therefore, spectral features, such as cutoff and bumps, can be expected at high energies. In this work we give a careful
study on the perspective of the electron spectrum beyond TeV. We
first examine our astrophysical source models on the latest leptonic data of AMS-02 to give a self-consistent picture. Then we focus on the discussion about the candidate sources which could be electron contributors above TeV.
Depending on the properties of the local sources (especially on the nature of Vela), DAMPE may detect interesting features in the electron spectrum above TeV in the future.
\end{abstract}

\maketitle

\section{Introduction}
The Alpha Magnetic Spectrometer (AMS-02) launched in May 2011 has
taken the measurement of cosmic ray (CR) leptonic spectra to a new level \cite{2013PhRvL.110n1102A}.
Unprecedentedly precise results have been released in the energy range from
$\sim1$~GeV to $\sim500$~GeV for intensities of both electron and positron. The
well-known electron/positron excess, which was uncovered by earlier satellite experiments such as the Payload for
Antimatter Matter Exploration and Light-Nuclei Astrophysics (PAMELA) \cite{2009Natur.458..607A,2011PhRvL.106t1101A} and the
Fermi Large Area Telescope (Fermi-LAT)
\cite{2009PhRvL.102r1101A,2010PhRvD..82i2004A}, and also the anterior
balloon-borne Advanced Thin Ionization Calorimeter (ATIC) experiment
\cite{2008Natur.456..362C}, has been confirmed by AMS-02. Sources which have the potential to
provide primary electron/positron pairs, e.g., pulsars and annihilation of dark
matter (DM), are thus involved into CR models to interpret those excesses.
However, some global fittings of AMS-02 leptonic spectra indicate that the electron spectrum
has a larger excess to the background than that of positron
\cite{2014PhLB..728..250F,2015PhLB..749..267L,2015PhRvD..91f3508L}. As the
contributions of $e^{\pm}$ pairs from exotic sources
like pulsars or DM are constrained by the positron spectrum, these sources seem not enough to explain the total electron excess.

A possible interpretation of the extra electron excess is the hardening in high
energy range of the electron spectrum of the supernova remnant (SNR) background,
which can be attributed to the fluctuation given by local discrete SNRs \cite{2015PhRvD..91f3508L}. Thus
the concept of dividing local SNRs and distant SNRs steps again into the
spotlight. This concept was first put forward by \citet{shen70} and improved by
later
works. In the model of \citet{1995PhRvD..52.3265A}, a nearby ($\leq100$~pc) and
relatively young
($\leq10^5$~yr) source and continuously distributed distant sources
($\geq1$~kpc) contribute separately to the electron spectrum; they also adopted
a
energy dependent diffusion coefficient in the propagation model. \citet{koba04}
went further on this scenario by using several real sources with known ages and
distances as local electron accelerators. After the publishing of AMS-02 data,
\citet{mauro14} fit all the leptonic data simultaneously, applying the method
similar to the works above when dealing with local and distant SNRs.  They derived
spectral index and normalization of injection spectrum of individual SNR from
radio observations, comparing with uniform injection spectral parameters of all
the sources adopted in Ref. \cite{koba04}.

In fact, there are alternative explanations toward the electron/positron excess,
such as improper propagation parameters used in previous works \cite{2016ApJ...824...16J}. In these cases,
exceptional consideration of local sources may not be necessary to keep
consistency with the AMS-02 data. However, the ground-based Cherenkov telescopes
like the High Energy Stereoscopic System (HESS) \cite{hess08,hess09} and the Very Energetic Radiation
Imaging Telescope Array System (VERITAS) \cite{veritas} seemed to detect a cut-off around
$\sim1$~TeV in the $e^-+e^+$ spectrum, which cannot be described by a continuous
distributed background. The Dark Matter Particle Explorer (DAMPE)
\cite{dampe} launched in December 2015 aims to measure electrons in the range
of 5~GeV$-$10~TeV with unprecedented energy resolution (1.5\% at 100 GeV). As nearby sources have the
potential to contribute to the
highest energy range covered by DAMPE, we can expect to see some spectral
features in future DAMPE results. In this case, separating local sources
from the continuous distribution would be inevitable.

Basing on previous works mentioned above, we perform a careful analysis of
the local SNRs and their parameters of injection spectra. We
assume that pulsars are extra positron sources, and perform global
fittings to the latest leptonic data of AMS-02 for several SNR parameter settings.
We show below that the choice of parameters of a particular SNR has a significant
influence on its contribution. Although the electron energy range
covered by AMS-02 is under TeV, fittings to the AMS-02 leptonic data provide a self-consistent
picture for the astrophysical source models. As the local sources accounting for the AMS-02
results may provide contribution to the TeV scale, the AMS-02 data could also constrain the properties of the
predicted $e^-+e^+$ spectrum above $\sim$ TeV.
Combining with the fitting results, we then discuss the parameters
of the local sources which have the potential to contribute to TeV and give
further predictions of the $e^-+e^+$ spectrum up to the energy range of
10~TeV, which can be measured by DAMPE.

This paper is organized as follows. In Sec. \ref{sec:method}, we describe our
calculation towards the injection and propagation of Galactic electrons and
positrons to get leptonic spectra. The results of global fittings to leptonic
data of AMS-02 and our further predictions to the electron spectrum in the TeV range are
presented in Sec.
\ref{sec:ams} and Sec. \ref{sec:tev}, respectively. We summarize our work in the
last section.

\section{Method}
\label{sec:method}
In this section, we introduce the semi-analytical solution to the
propagation equation in the first part. Then we discuss the possibly that
SNRs are the most important sources of high energy CR electrons. Discussions of other
sources, including pulsars and secondary electrons/positrons, are given in the later subsections.

\subsection{Propagation of Cosmic Rays}
\label{subsec:ppg}
The propagation of CR electrons in the Galaxy can be described by the diffusion
equation with additional consideration of energy loss during their journey,
which may be written as
\begin{equation}
 \frac{dN}{dt} - \nabla(D\nabla N) - \frac{\partial}{\partial E}(bE) = Q \,,
 \label{eq:diff}
\end{equation}
where $N$ is the number density of particles, $D$ denotes the diffusion
coefficient, $b$ denotes the energy-loss rate
and $Q$ is the CR source function. Galactic convection and
diffusive reacceleration
are not taken into account here, since they have little effect above
10~GeV \cite{dela10}. We treat
the propagation zone
of CRs as a cylindrical slab, with radius of 20 kpc \cite{dela10} and a half thickness $z_h$.
The diffusion coefficient $D$ depends on the energy of CRs, which has the form
$D(E)=\beta D_0{(R/\rm 1\,GV)}^{\delta}$, where $D_0$ and $\delta$ are both
constants, $\beta$ is the velocity of particles in the unit of light speed and
$R$ is the rigidity of CRs. To give a constraint of major
propagation parameters---($D_0$,$\delta$,$z_h$), Boron-to-Carbon ratio (B/C) is
widely used. Unstable-to-stable beryllium (${}^{10}{\rm Be}/{}^{9}{\rm Be}$) is
also helpful to constrain CR propagation. We adopt the B/C data of ACE
\cite{ACEbc}, AMS-02 \cite{AMSbc}, ${}^{10}{\rm Be}/{}^{9}{\rm Be}$ data of
Ulysses \cite{ulyssesbe}, ACE \cite{ACEbe}, Voyager \cite{voyagerbe}, IMP
\cite{1988SSRv...46..205S}, ISEE-3 \cite{1988SSRv...46..205S}, ISOMAX
\cite{ISOMAXbe}, and embed the CR propagation code in the Markov Chain Monte
Carlo (MCMC) sampler to acquire best fitted propagation parameters (this work is
in preparation). We adopt $D_0=2.12\times10^{28}\,\rm cm^2 s^{-1}$,
$\delta=0.548$, $z_h=3.8$~kpc in this work.

\begin{figure}
\centering
\includegraphics[width=0.65\textwidth, angle=270]{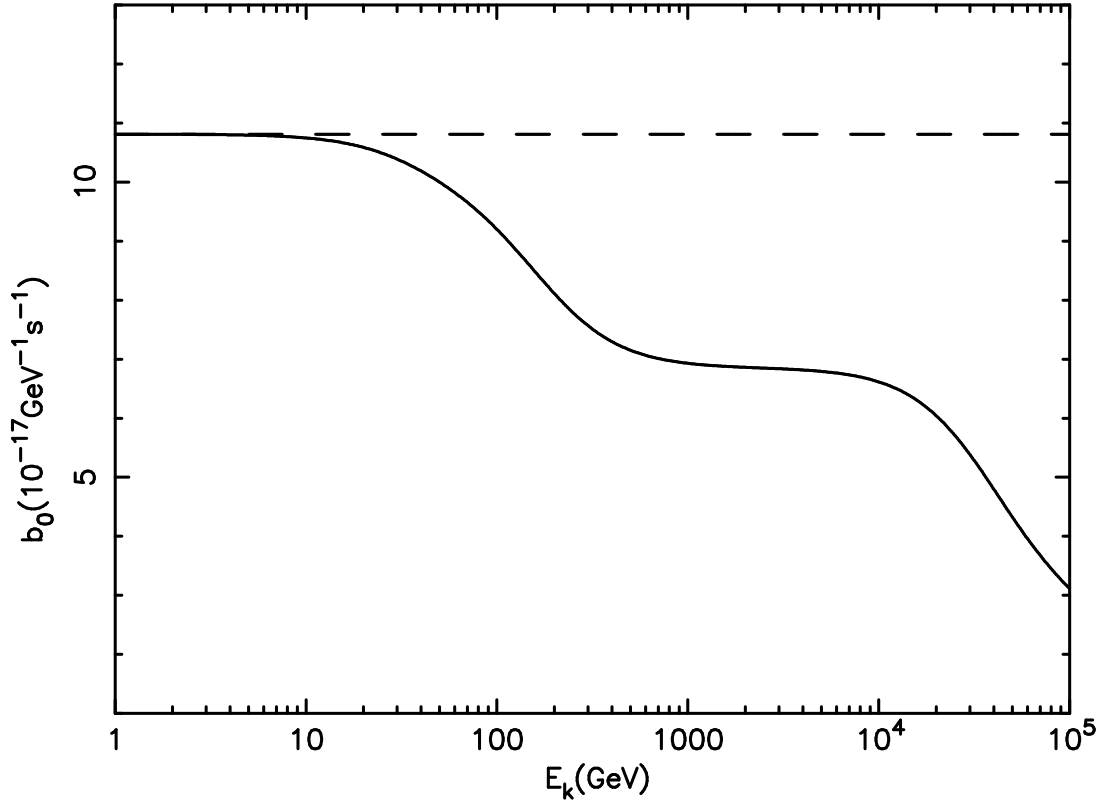}
\caption{Energy-loss coefficient $b_0$ as a function of $e^{\pm}$ energy. The
dashed line stands for the case of Thomson approximation and the solid line
describes the drops of $b_0$ due to Klein-Nishina correction.}
\label{fig:b0}
\end{figure}

Positrons and electrons with energy higher than 10~GeV suffer from energy
loss during their propagation mainly by synchrotron radiation in the Galactic
magnetic field and inverse Compton radiation in the interstellar radiation
field consisting of stellar radiation, reemited infrared radiation from
dust, and cosmic microwave background (CMB). We set the interstellar magnetic
field in the Galaxy to be $1~\mu$G to get the synchrotron term \cite{dela10}. For the inverse
Compton process, if we use the cross section for Thomson scattering, the
energy-loss rate $b(E)$ has a quadratic dependence on energy:
\begin{equation}
 \frac{dE}{dt} = -b_0E^2\,,
 \label{eq:b}
\end{equation}
where $b_0$ is a constant. However, when the energy comes higher, a
relativistic correction to the cross section, namely a Klein-Nishina cross
section, is needed. Here we adopt the description of Ref.~\cite{schli10} which
gives a reconcilable approximation between Thomson limit and Klein-Nishina
limit. The temperature and energy density of radiation field components are:
20000~K and $0.09\,\rm{eV\,cm^{-3}}$ for type B stars, 5000~K and
$0.3\,\rm{eV\,cm^{-3}}$ for type G-K stars, 20~K and $0.4\,\rm{eV\,cm^{-3}}$
for infrared dust, and 2.7~K and $0.25\,\rm{eV\,cm^{-3}}$ for CMB. In this case,
$b_0$ is not a constant anymore but decreases with the energy as shown
in Fig.~\ref{fig:b0}. We can find from Fig.~\ref{fig:b0} that the
relativistic correction becomes important as long as energies of electrons/positrons
are higher than 10~GeV. We still use the symbol $b_0$, which has the
connotation of $b_0(E)$, in our later work for convenience.

\begin{figure}
\centering
\includegraphics[width=0.65\textwidth, angle=270]{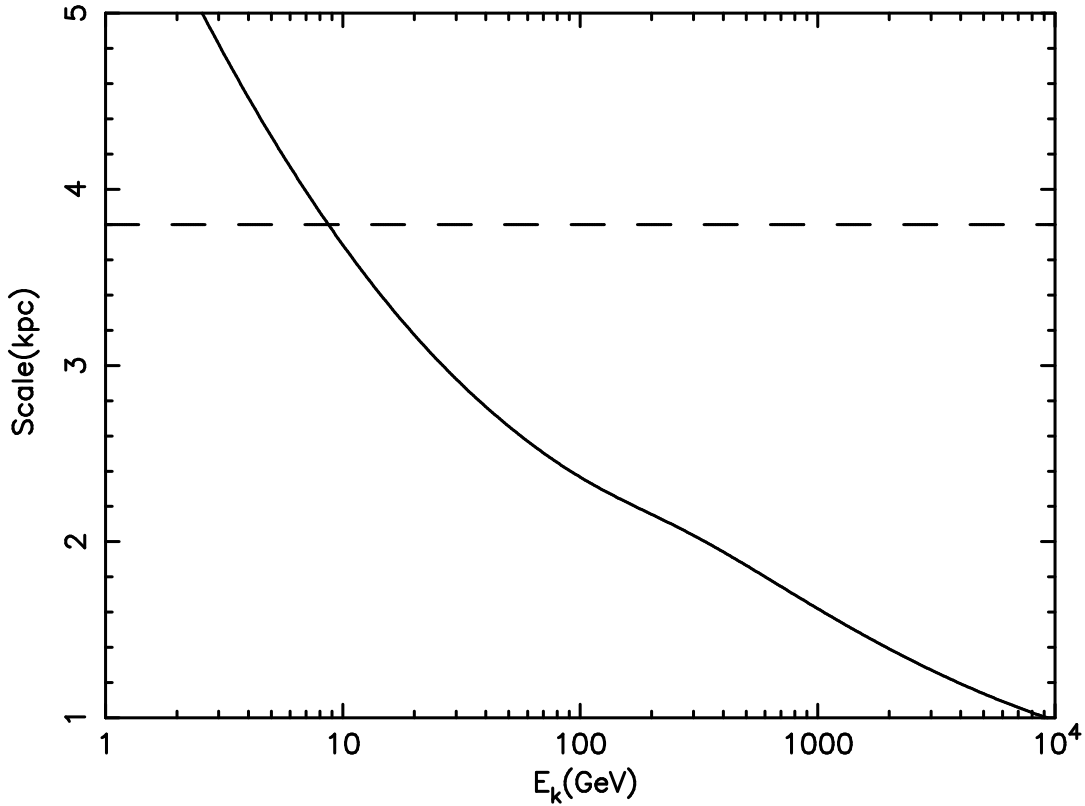}
\caption{Comparision between diffusion distance of CRs and half-thickness of
propagation zone. The solid line represents the diffusion scale
$\lambda\sim2\sqrt{D(E)t}$ while the dashed line marks $z_h$.  }
\label{fig:scale}
\end{figure}

Cylindrical coordinate is applied in our work to describe the disc-like
geometry of the propagation zone, and the location of the Earth is set to be
zero. For a point source with burst-like injection, the source function can be
written as
\begin{equation}
 Q(E,t,r,z)=Q(E)\delta (t-t_s)\delta (r-r_s) \delta (z-z_s)\,,
 \label{eq:Q4D}
\end{equation}
where $Q(E)$ represents the energy distribution of injection, $t_s$ is the time
of CR injection, $r_s$ and $z_s$ are radial and verticle location of the
source, respectively. As long as Galactic CR sources are mostly distributed in
a much thinner vertical scale comparing with $z_h$ and so does the solar
system, we assume $z_s=0$ in our work for all the sources. Then the
time-dependent Eq.(\ref{eq:diff}) can be solved semi-analytically with the help
of
Green's function $G(E,t,r\gets E_s,t_s,r_s)$ working in the Fourier space. We
follow the Green's function used in Ref.~\cite{koba04}:
\begin{equation}
\begin{aligned}
 G(E,t,r\gets E_s,t_s,r_s) = &\,\,\delta(E_s-E_0)\,\frac{b(E_0)}{b(E)}\,
\frac{1}{\pi\lambda^2}\,{\rm
exp}\left[-\frac{(r-r_s)^2}{{\lambda}^2}\right] \\
                             &\,\,\times\sum_{n=0}^{\infty}\frac{1}{z_h}\,{\rm exp}\left(-\frac{{\lambda}^2k^2_n}{4}\right)\,,
\end{aligned}
\end{equation}
where $k_n=(2n+1)\pi/(2z_h)$, $E_0=E/[1-b_0E(t-t_s)]$, and
\begin{equation}
 \lambda\equiv 2\left(\int_{E}^{E_0}\frac{D(E')dE'}{b(E')}\right)^{1/2}
 \label{eq:lambda}
\end{equation}
is the diffusion distance for particles with initial energy $E_0$ and final
energy $E$. The solution of Eq.~(\ref{eq:diff}) has the form of
$G(E,t,r\gets E_s,t_s,r_s)Q(E_s,t_s,r_s)$, so the observed CRs contributed by a
source with distance $r$ and age $t$ should be expressed as
\begin{equation}
\begin{aligned}
 N(E,t,r) = &\,\, \frac{1}{\pi\lambda^2}\,{(1-b_0Et)}^{-2}\,{\rm
exp}\left(-\frac{r^2}{\lambda^2}\right)\,Q
 \left(\frac{E}{1-b_0Et}\right)\\
            &\,\,\times \sum_{n=0}^{\infty}\frac{1}{z_h}\,{\rm exp}\left(-\frac{{\lambda}^2k^2_n}{4}\right)\,.
\end{aligned}
\label{eq:N}
\end{equation}
However, considering the efficiency of doing numerical calculation,
Eq.~(\ref{eq:N}) may not be a good expression since we need to include more
terms in higher energy range to guarantee precision of the calculation. Thus a
spherically symmetric time-dependent solution to Eq.~(\ref{eq:diff}), which has a
form of \cite{ginz76,maly09}
\begin{equation}
 N(E,t,r) = \frac{1}{(\pi\lambda^2)^{3/2}}\,{(1-b_0Et)}^{-2}\,{\rm
exp}\left(-\frac{r^2}{\lambda^2}\right)\,Q
 \left(\frac{E}{1-b_0Et}\right) ,
 \label{eq:Nsph}
\end{equation}
can be treated as a substitute of Eq.~(\ref{eq:N}). Given the disc-like geometry
of propagation zone, Eq.~(\ref{eq:Nsph}) is valid when $\lambda \ll z_h$.
Assuming $E$ is close enough to $E_0$, the diffusion distance defined in
Eq.~(\ref{eq:lambda}) can be approximated by $2\sqrt{D(E)t}$.
As shown in Fig.~\ref{fig:scale}, the condition $\lambda < z_h$ is always satisfied
as long as $E>10\,$GeV, and the difference increases in higher energy. In fact,
we find the error is less than 1\% for $E>10\,$GeV after calculating ratios
between these two expressions.

\subsection{SNRs as Electron Source}

\subsubsection{Local SNRs and SNR Background}
SNRs are believed to be the main astrophysical sources of primary Galactic CRs,
such as nucleus and electrons. Particles can be boosted to very high energy
through diffusive shock wave acceleration in SNR. However, among the
accelerated particles, electrons/positrons undergo significant energy loss
during their propagation through electromagnetic radiation. Eq.~(\ref{eq:b})
indicates the life time of an electron is roughly $1/(b_0E)$, thus electrons
with higher energy become inactive faster. Since $\delta<1$, the diffusion
distance $2\sqrt{D(E)t}$ has an inverse relationship with electron energy. For
example, electrons in $\sim\,$TeV fade within a radius of roughly 1~kpc from
their source. Thus high energy part of electron spectrum can only be contributed
by several local sources, and a simple continuous source distribution is no
longer valid due to the spectral fluctuations induced by those few sources. Then
it is important to separate local discrete sources from distant sources in
calculation, as first proposed by \citet{shen70}. We
treat SNRs within 1 kpc as local sources and farther sources as
background contributors of electrons \cite{koba04}. The intensity of electrons
from a SNR nearby can be simply expressed by
\begin{equation}
 I(E) = \frac{c}{4\pi}N(E,t,r)\,.
 \label{eq:local}
\end{equation}

In order to obtain distant components, we calculate the electron spectrum produced by a
smooth distribution of SNRs in whole range of distance and age first, and then
subtract the local components in continuous form. We set supernova explosion
rate to be $f=4~{\rm century^{-1}\,galaxy^{-1}}$ \cite{dela10}. Since about
2/3 of
supernovae are expected to be type II supernova, we can use the
population of pulsars to describe the spatial distribution of SNRs. Here we
choose the Galactic distribution of pulsars given by Ref.~\cite{l04} as
\begin{equation}
 \rho (R) \sim R^n\,{\rm exp}\left(-\frac{R}{\sigma}\right)\,,
\end{equation}
where $n=2.35$,$ \sigma=1.528$~kpc and $R$ is the distance to the Galactic
center. Note that the zero point of this distribution
is the Galactic center, rather than the solar system used in our work. Thus
$\rho$ should also depend on the azimuth angle $\varphi$. Here we aim to find
how local sources create spectrum features in TeV range. As the diffusion distance of 1~kpc corresponds to a electron cooling time
of roughly $3\times 10^5\,$yr, sources older than this age are treated as background SNRs in our calculation. Finally we get the electron
spectrum of background component:
\begin{equation}
\begin{aligned}
 I(E) = &\,\,\frac{c}{4\pi}\,\left(\int_{0}^{\frac{1}{b_0E}}dt\int_{0}^{\infty}dr\int_{0}^{2\pi}d\varphi
 -\int_{0}^{t_m}dt\int_{0}^{r_m}dr\int_{0}^{2\pi}d\varphi\right) \\
 &\,\,\times f\,\rho(r,\varphi)\,N(E,t,r)\,r\,,
\end{aligned}
 \label{eq:bkg}
\end{equation}
where $\rho(r,\varphi)$ is the normalized distribution, $r_m=1$ kpc and
$t_m=3\times 10^5$ yr for the case $1/(b_0E)>3\times 10^5$ yr (otherwise $t_m$
takes $1/(b_0E)$). Indeed, full propagation equation with consideration of
convection and
reacceleration may be solved with public numerical tool GALPROP \cite{galp}
which can give
a
more accurate result. Nevertheless, the spatial zero point used in GALPROP is
the center of the Galaxy, we may need to do some troublesome work to divide
local sources from the background. That's the reason we choose an semi-analytic
treatment toward Eq.~(\ref{eq:diff}).

\begin{table}
    \begin{tabular}{| c| c| c| c| c |c| c| c |}
\hline
Source    &Other Name    &$B_r^{\rm 1GHz}$[Jy] &$\alpha _r$ &Size[arcmin]
&r[kpc]  &t[kyr]  & Ref.  \\ \hline
G065.3+05.7      &-             &52            &0.58           &$310\times240$
&0.9      &26
&\cite{green,1996ApJ...458..257G,2002AA...388..355M,2009AA...503..827X}    \\
\hline
G074.0--08.5     &Cygnus Loop   &175           &0.4            &$230\times160$
&0.54     &10    &\cite{green,2005AJ....129.2268B,2006AA...447..937S}   \\
\hline
G114.3+00.3      &-             &6.4           &0.49           &$90\times55$
&0.7      &7.7
&\cite{green,2013IJMPS..23...82H,2006AA...457.1081K,2004ApJ...616..247Y}   \\
\hline
G127.1+00.5      &R5            &12            &0.43           &45       &1
  &$[20,30]$     &
\cite{green,2013IJMPS..23...82H,1989AA...219..303J,2006AA...457.1081K,
2006AA...451..251L} \\ \hline
G156.2+05.7      &-             &5             &0.53           &110      &1.0
  &$[15,26]$
&\cite{green,2009PASJ...61S.155K,2006AA...457.1081K,1992AA...256..214R,
2007AA...470..969X,2000bbxs.conf..567Y}\\ \hline
G160.9+02.6      &HB9           &88            &0.59           &$140\times120$
&0.8      &$[4,7]$
&\cite{green,2013IJMPS..23...82H,2006AA...457.1081K,2007AA...461.1013L,
2003AA...408..961R}\\ \hline
G203.0+12.0      &Monogem Ring  &-             &-              &-        &0.3
 &86        &\cite{plucinsky96,plucinsky09} \\ \hline
G263.9--03.3     &Vela YZ       &varies        &varies         &255      &0.29
  &11.3
&\cite{green,alvarez01,2001ApJ...561..930C,1999ApJ...515L..25C,
2008ApJ...676.1064M}\\ \hline
G266.2--01.2     &Vela Jr.       &50            &0.3            &120      &0.75
  &$[1.7,4.3]$
&\cite{green,1998Natur.396..141A,1998Natur.396..142I,2008ApJ...678L..35K,
2005MNRAS.356..969R}\\ \hline
G328.3+17.6      &Loop I (NPS)  &-             &-              &-        &0.1
 &200      &\cite{bingham67,egger95}\\ \hline
G347.3--00.5     &RXJ1713.7-3946&4             &0.3            &$65\times55$
&1        &1.6     &\cite{green,2004ApJ...602..271L,2009MNRAS.392..240M}
\\  \hline
\end{tabular}
\caption{Basic parameters for local SNRs within 1~kpc. The columns show the
Green catalog name, the association name, the radio flux $B_r^{\rm 1GHz}$, the
radio spectrum index $\alpha _r$, the distance $r$ and the SNR age $t$. Note
that $t$ is the observed age, the actual age should be $T=t+r/c$ .}
 \label{tab1}
\end{table}

\subsubsection{Parameters of SNRs}
\label{subsubsec:snr_para}
For SNRs, particles are accelerated by shock acceleration mechanism, or
say, the first order Fermi acceleration. The energy spectrum produced by
Fermi acceleration is thought to be a power law form. Taking energy loss
and escape of particles into account, the emergent spectrum of SNR can be
described by a power law form with an exponential cutoff:
\begin{equation}
 Q(E)=Q_0{(E/{\rm 1\,GeV})}^{-\gamma}{\rm exp}(-E/E_c)\,,
 \label{eq:Q1D}
\end{equation}
where $Q_0$ is the normalization of the injection spectrum. Note that
distant SNRs are treated as a continuous distribution. Thus we assume that they share
common $Q_0$ and $\gamma$, and set these two parameters to be free in
following fittings. The energy cut-off is fixed at 20~TeV for background
sources. For local SNRs, we attempt to investigate their
parameters individually. Table \ref{tab1} lists objects locating within 1~kpc
included by the Green catalog of SNRs \cite{green} (of course those with a
measured distance) with two additional sources Monogem
Ring and Loop I.

Several SNRs have gone through multi-wavelength
measurements, from radio to $\gamma$-ray bands, which are helpful to estimate
parameters in Eq.~(\ref{eq:Q1D}). These sources include HB9 (radio and $\gamma$-ray),
Vela Jr. (radio, X-ray and $\gamma$-ray), RX J1713.7-3496 (radio, X-ray and
$\gamma$-ray), and Cygnus loop (radio and $\gamma$-ray). Note that we just pick
observations with available data. Radio and X-ray emissions are produced
by the synchrotron radiation of relativistic electrons accelerated in the SNR,
while $\gamma$-ray emissions could have either a leptonic origin or a hadronic
origin, which are generated through scattering of background photons by
relativistic electrons or through $\pi^0$ decay resulting from the collision of
accelerated protons with ions
in the background plasma. As we are interested in the electron
spectrum, a purely leptonic model is the priority in our fitting. If the leptonic
fitting fails, this model is replaced by a hybrid model, where
contributions from electrons and protons are comparable in $\gamma$-ray
spectrum. Purely hadronic model would not be
discussed in our work.

For the leptonic model, the energy spectrum of
accelerated electrons still has the form of Eq.~(\ref{eq:Q1D}), while we remark
the parameters in those expression as $Q_{0,e}$, $\gamma_e$ and $E_{c,e}$, to
distinguish from the case of proton. The background
radiation fields consist of interstellar infrared radiation, optical radiation
and CMB. The contributions to the inverse Compton $\gamma$-ray spectrum from the interstellar radiation field (ISRF) are more important
than that from CMB \cite{yuan11}. Here we
adopt the ISRF model given by Ref.~\cite{porter06}. In the hybrid model, the energy
spectrum of protons has the same form as that of electrons, where $Q_{0,p}$,
$\gamma_p$ and $E_{c,p}$ are corresponding normalization parameter, spectral
index and cut-off energy of protons. Assuming charged particles share the same
acceleration mechanism, the spectral index of protons could be identical to
that of electrons \cite{petro04,yuan11}. Thus the four free parameters in leptonic model are
$Q_{0,e}$, $\gamma_e$, $E_{c,e}$ and magnetic field $B$. For the case of hybrid
model, there are three additional parameters, $Q_{0,p}$, $E_{c,p}$ and number
density of target proton $N_{\rm H}$.

\begin{figure}
\centering
\includegraphics[width=0.4\textwidth]{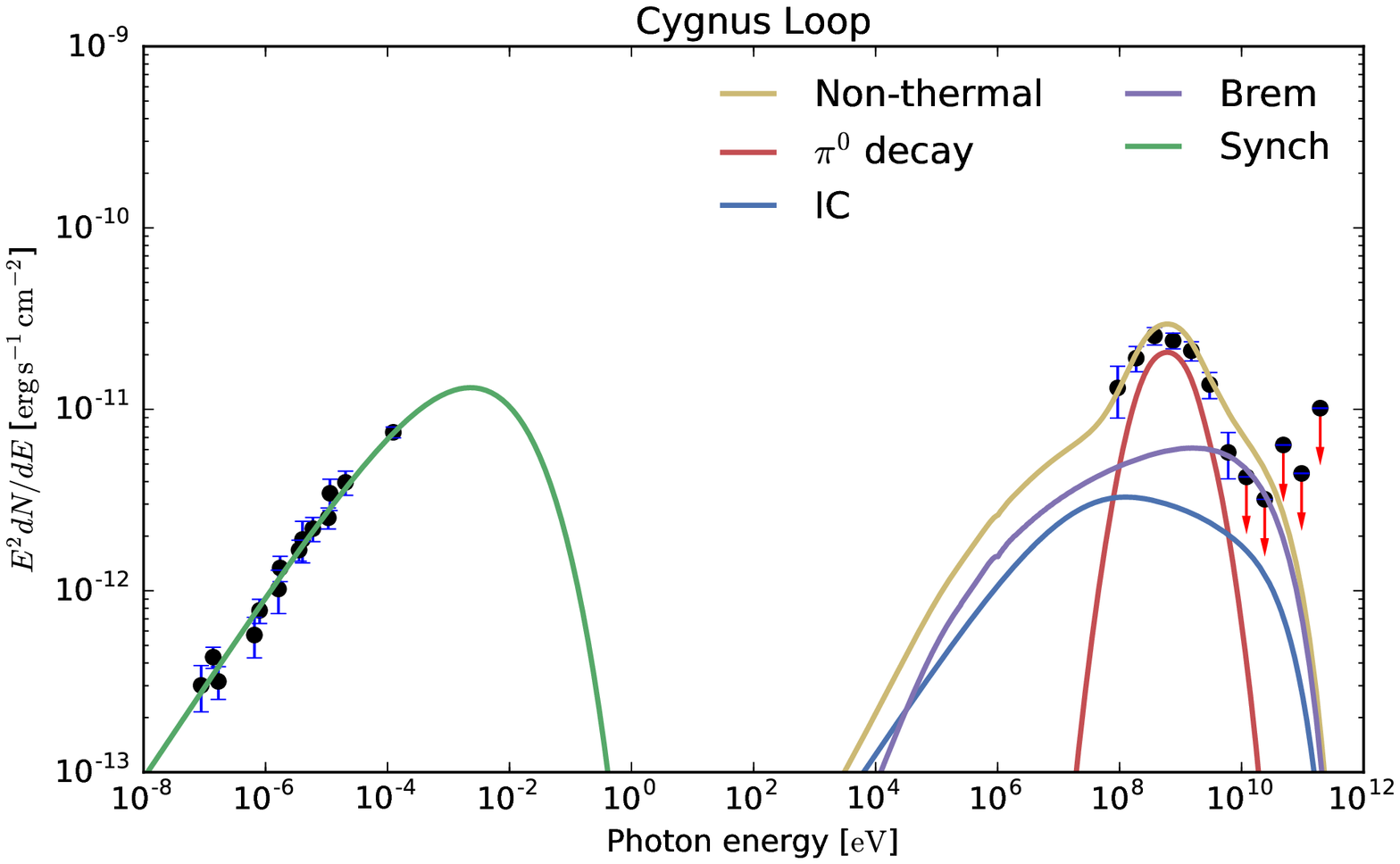}
\includegraphics[width=0.4\textwidth]{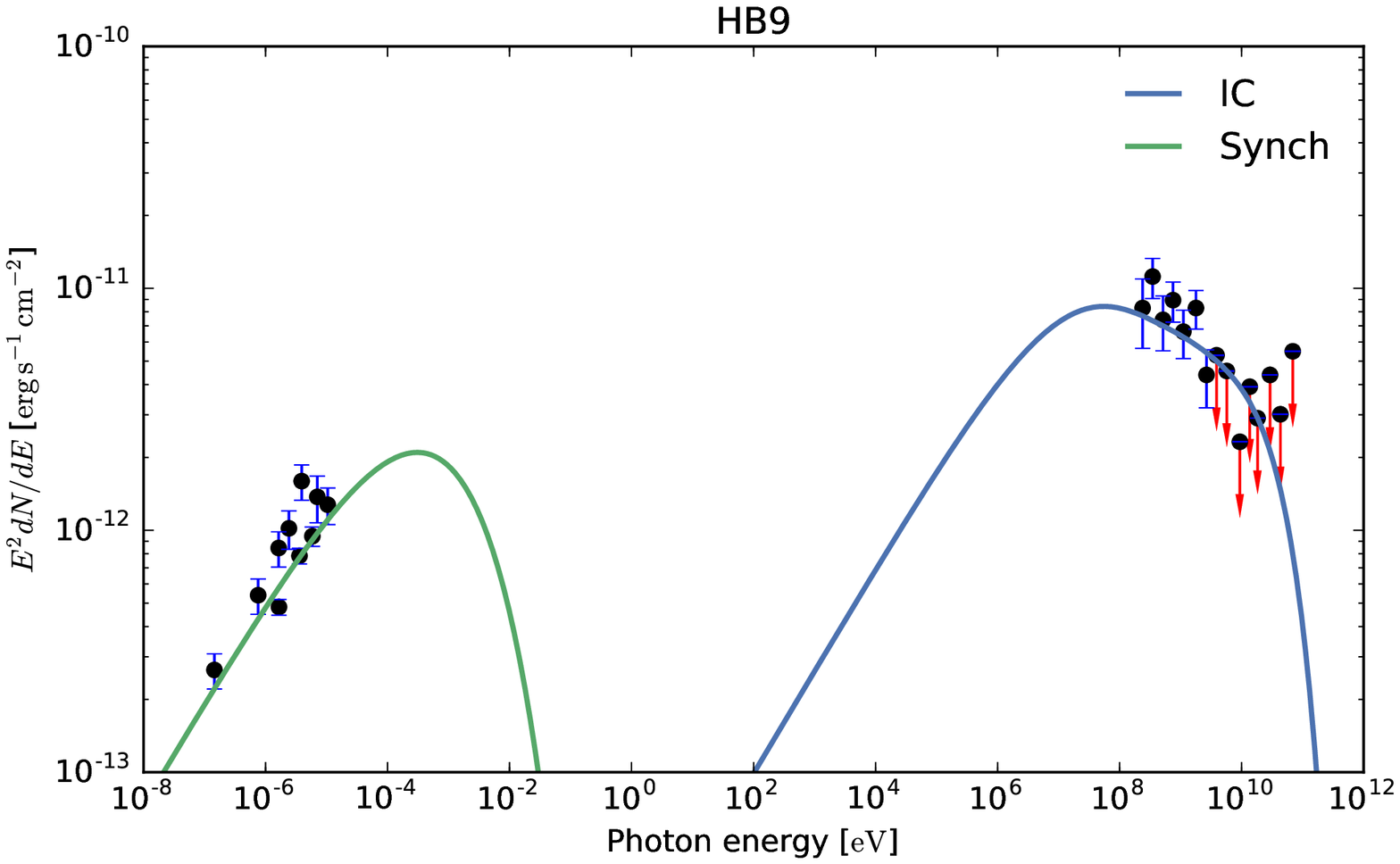}
\includegraphics[width=0.4\textwidth]{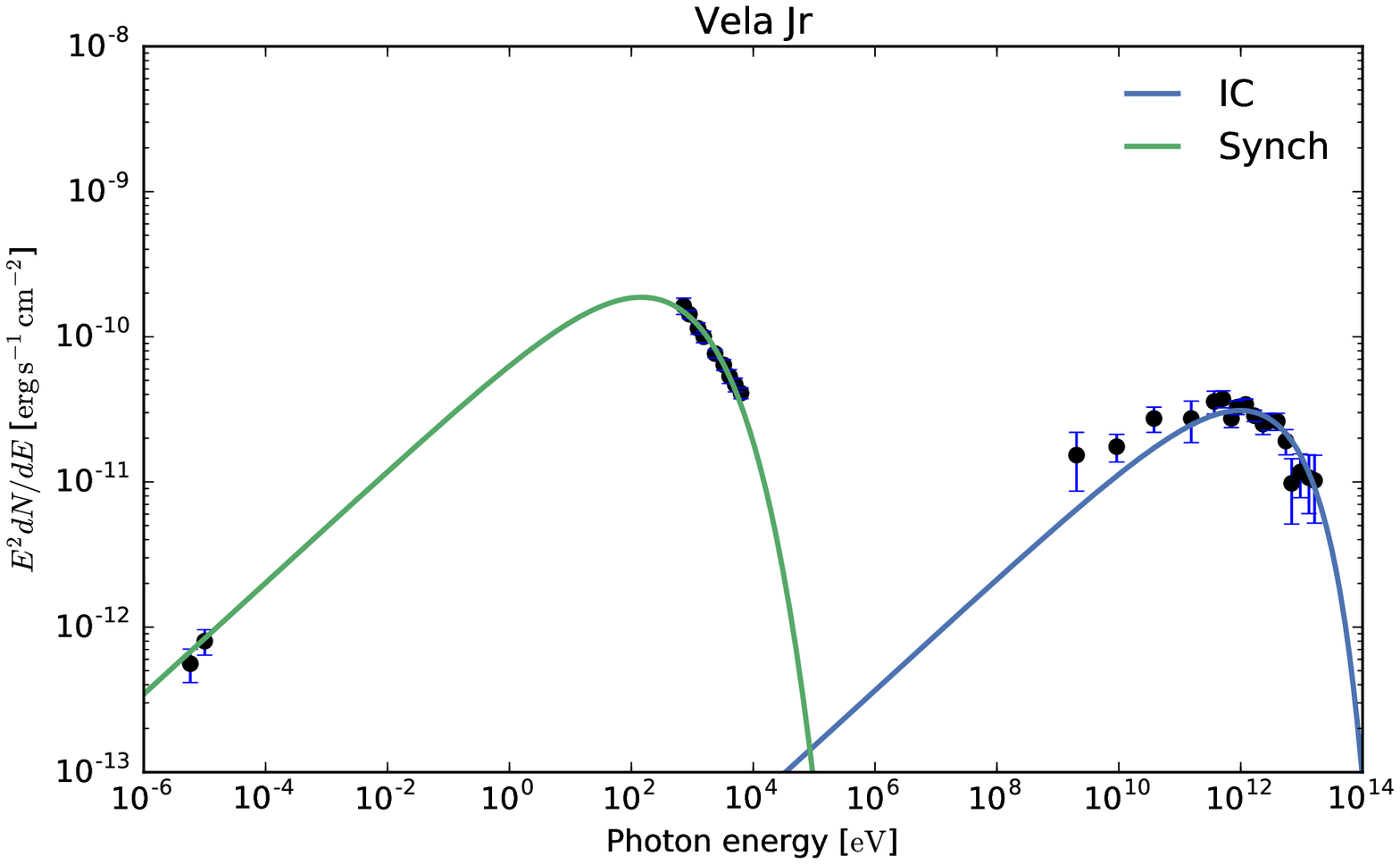}
\includegraphics[width=0.4\textwidth]{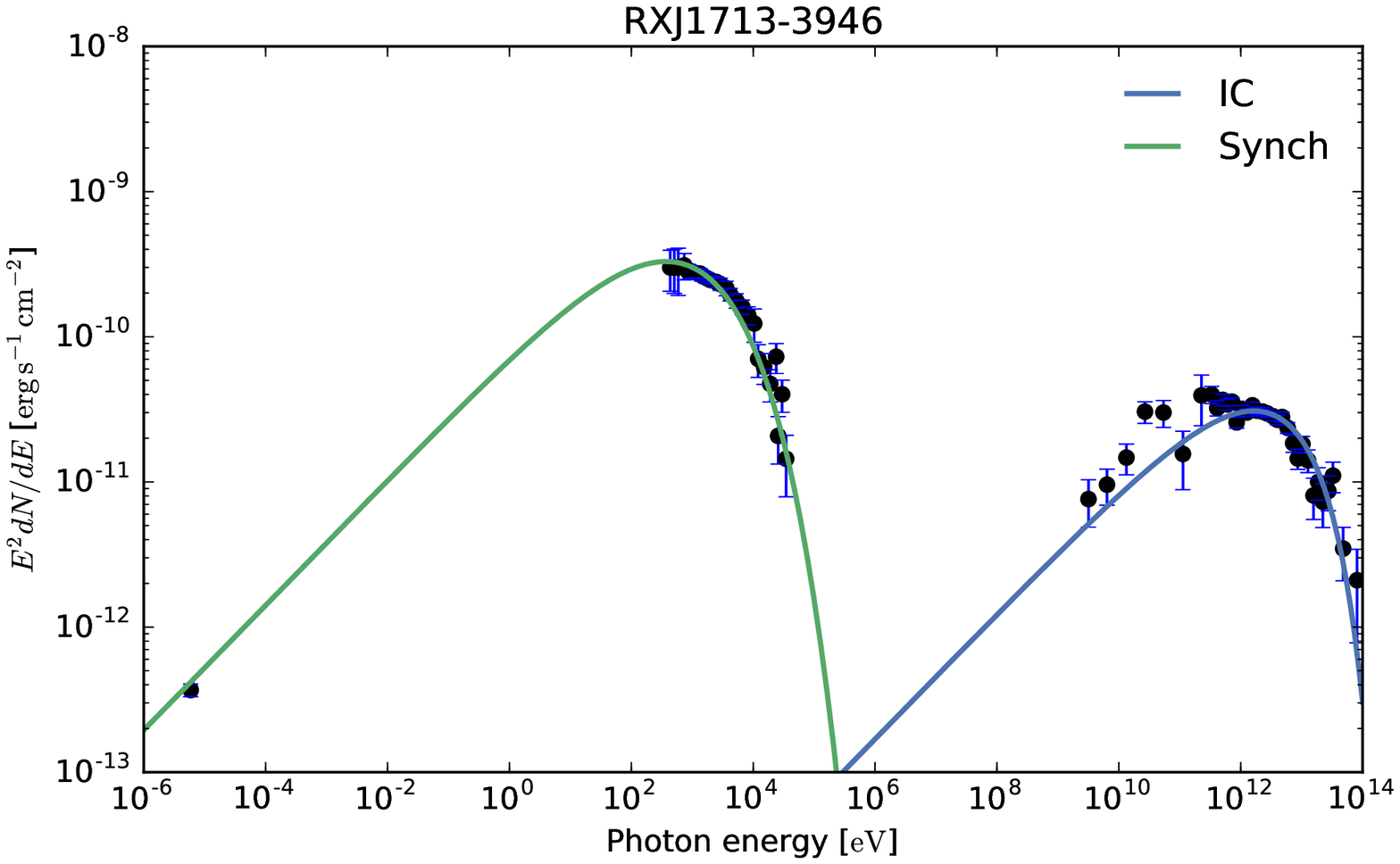}
\caption{Energy spectrum distribution of the sources Cygnus Loop, HB9, Vela Jr.
and RX J1713.7-3946. The components
 are: synchrotron (green line), IC(blue line), non-thermal bremsstrahlung
(purple line), neutral pion decay
(red line). Total $\gamma$-ray emission for hybrid scenario is marked by yellow
line. References for the fitted data points are given in Table \ref{tab2}.}
\label{fig:broadband}
\end{figure}

We apply naima, a python-based package, for computing non-thermal radiation
processes and fitting the spectral energy distributions \cite{naima}. It uses
Markov-Chain Monte Carlo emcee sampling \cite{emcee} to find
best-fit parameters of physical radiation models
and thus determines the radiation mechanism behind the observed
emission. The parameters of the best-fit models are compiled in Table \ref{tab2}
and the best-fit spectra are shown in Fig.~{\ref{fig:broadband}}.

Different from four SNRs mentioned above, we need other approaches to
determine parameters for other SNRs without precise high energy $\gamma$-ray
observations. Assuming the radio emissions of SNRs are entirely induced by
synchrotron emissions of electrons, \citet{mauro14} provided an estimation
method of $Q_0$ as
\begin{equation}
         Q_0 = 1.2\times10^{47}\; {\rm GeV}^{-1}\; (0.79)^{\gamma}
\left(\frac{r}{\rm{kpc}}\right)^2
 \left(\frac{B}{100 \mu
\rm{G}}\right)^{-(\gamma+1)/2}
	\left(\frac{B_r^{\rm{1GHz}}}{\rm{Jy}}\right)\, ,
 \label{eq:Q0}
\end{equation}
which relies on spectral index $\gamma$, source distance $r$, magnetic field
$B$ in SNR and the radio flux at 1~GHz $B^{1\rm{GHz}}_{r}$.
For the synchrotron
emission of electron system, the electron spectral index can be derived from the photon
spectral index. Thus we have $\gamma=2\alpha_r+1$, where $\alpha_r$ is the radio
spectral index. We list $B^{1\rm{GHz}}_{r}$ and $\alpha_r$ in Table \ref{tab1}. Another
critical parameter $B$ can only be approximately estimated by several methods,
such as Zeeman effect, Faraday rotation, and minimum energy or equipartition. The
equipartition or minimum energy method calculates magnetic field only depending on the
radio synchrotron emission of source \cite{beck05}. It is a useful tool if no other data
of the source is available. According to the result of \citet{arbutina11}, we can
obtain the magnetic field of SNR as
\begin{equation}
 \begin{split}
   B[\rm G] \approx &\,\,
   \left[
 6.286\times 10^{(9\gamma
-79)/2}\,\frac{\gamma+1}{\gamma-1}\,\frac{\Gamma(\frac{3-\gamma}{2}
)\Gamma(\frac{\gamma-2}{2})\Gamma(\frac{\gamma+7}{4})}{\Gamma(\frac{\gamma+5}{4
} )}(m_ec^2)^{2-\gamma} \right. \\
& \left.\,\,\times
\frac{(2c_1)^{(1-\gamma)/2}}{c_5}\,(1+\kappa)\,\frac{B_r^{\rm
1GHz}[{\rm Jy}]}{f\,r[{\rm kpc}]\,\theta[{\rm
arcmin}]^3}
 \right]^{2/(\gamma+5)}\,,
\end{split}
\label{eq:B}
\end{equation}
where $m_ec^2$ is rest energy of electron, $c_1$ and
$c_5=c_3\Gamma{(\frac{3\gamma-1}{12})}\Gamma{(\frac{3\gamma+19}{12})}
/(\gamma+1)$ are defined in Ref.~\cite{pach70},  $f$ is the volume filling factor of radio
emission, $\theta$ is the angular radius of SNR which can be found in Green
catalog. In Eq. (\ref{eq:B}), $\kappa$ is a parameter depending on $\gamma$ and
ion abundances of SNR. We adopt a typical value $f=0.25$ and simple ISM
abundance H:He=10:1 in our estimation. Note that this minimum energy method
is applicable only for mature SNRs with $\alpha>0.5$. For G114.3+00.3 and
G127.1+00.5, we take a typical value of $B=30\,\rm{\mu G}$. Finally, We follow
the method proposed by \citet{yamazaki06} to give an
estimation to $E_c$. When the age of a SNR meets $t\gtrsim10^3\,$yr,
synchrotron loss restricts the maximum energy of electrons. Thus we have
\begin{equation}
E_c = 14\,{\rm TeV}\,h^{-1/2}\,v_{s,8}\,(B/10{\rm \mu G})^{-1/2}\,,
\label{eq:Ec}
\end{equation}
where $v_{s,8}$ is
the shock velocity in unit of $10^8\,{\rm cm\,s^{-1}}$ which depends on
evolution of SNR. See details in Ref.~\cite{yamazaki06} for the definition of
$h$ and here we take it to be unit.

\begin{table}
\centering
\begin{center}
\begin{tabular}{|c|c|c|c|c|c|c|}
  \hline
  Source       &$B[\rm{\mu G}]$  &$Q_{0,e}[10^{49}{\rm GeV^{-1}}]$ &$\gamma _e$
&$E_{c,e}[\rm{TeV}]$  &$W_e[10^{48}\rm{erg}]$  &Ref. \\  \hline
      G065.3+05.7  &10.6        &10.5          &2.16         &7.2         &0.77
  &- \\  \hline
      G074.0--08.5 &9.7         &10            &1.99         &0.072       &0.63
&\cite{2011ApJ...741...44K,2015arXiv150203053R,2004AA...426..909U}\\ \hline
      G114.3+00.3  &30          &0.14          &1.98         &9.3         &0.02
  &-  \\ \hline
      G127.1+00.5  &30          &0.52          &1.86         &4.6         &0.12
  &-  \\ \hline
      G156.2+05.7  &13          &0.83          &2.06         &8.4         &0.09
  &-   \\ \hline
      G160.9+02.6  &2.3         &130           &2.15         &0.065       &5.66
 &\cite{2014MNRAS.444..860A,1982JApA....3..207D,2007AA...461.1013L,
2003AA...408..961R} \\ \hline
      G266.2--01.2 &8.9         &9.6           &2.23         &25.4        &0.59
  &\cite{2007ApJ...661..236A,2000AA...364..732D,tanaka11}  \\ \hline
      G347.3--00.5 &11.9        &5.7           &2.14         &31          &0.48

&\cite{2011ApJ...734...28A,2011AA...531A..81H,2009AA...505..157A,
2015AA...577A..12F,2008ApJ...685..988T} \\ \hline
\end{tabular}
\caption{ The electron spectrum parameters for our sample. The parameters of
HB9, Vela Jr. and RXJ1713.7-3946
are obtained by fitting the multi-wavelength emission by leptonic model. The
hybrid model give better fitting for Cygnus loop,
and the extra parameter are $Q_{0,p}=1.5\times10^{52}\,{\rm GeV}^{-1}$,
$E_{c,p}=12\,\rm{GeV}$, $N_H=0.46\,\rm{cm^{-3}}$.
Magnetic field strength of  G065.3+05.7 and G156.2+05.7 are deduced from
Eq.~(\ref{eq:B}), and we adopt 30~$\mu G$ for G114.3+00.3 and G127.1+00.5.
For those sources $Q_{0,e}$ and $E_{c,e}$ are derived from Eq.~(\ref{eq:Q0})
and Eq.~(\ref{eq:Ec}) respectively, while $\gamma_e$ is given by the relation
$\gamma=2\alpha_r+1$. References for the multi-wavelength fitting data are given
in the last column.}
\label{tab2}
\end{center}
\end{table}

So far we have estimated parameters for all the sources in Table \ref{tab1}
except for Vela(XYZ), Monogem Ring and Loop I. The results are listed in Table
\ref{tab2}. Vela(XYZ) is generally believed to be an important local
source of CRs. We discuss it in detail in the next section. For Monogem ring and Loop I, they are classified as possible or probable SNRs by
Green \cite{green} and are not included in the catalog due to the lack of understanding to
them. We treat them as potential sources of electrons in the next
section and set their parameters to be free. Note that all
those estimations of SNR
parameters rely on the electromagnetic radiations of SNR. These emissions are
not produced by electrons observed today but some 'younger' ones in SNRs. This
implies that even if observations of photon emission are precise enough, the
derived electron parameters may be different from those of injected electrons due to the
variation of parameters along with the evolution of SNRs. Therefore the aim of our
estimation is to give some reference values for electron injection parameters, but far from to 'determine' them.

\subsection{PWN as Electron and Positron Source}
Pulsars are known to be the most important astrophysical sources of high energy
electron/positron pairs in the Galaxy \cite{torre13}. They produce relativistic wind carrying
charged particles at the cost of their spin-down energy \cite{rees74}. Since pulsar is formed
in SN explosion, it is initially surrounded by its companion SNR. When the
relativistic pulsar wind impacts on the cold SN ejecta which expands with a
slower velocity, a termination shock is formed besides a forward shock. The termination shock propagates
inward and reaches the radius where the outward pressure of pulsar wind balances
the internal pressure of the shocked bubble (see Ref.~\cite{pwn06} and
references therein). The shocked region is dubbed the
pulsar wind nebula (PWN). After particles inter the PWN, they are confined
by the magnetic field here for a long period, until the crushing of the PWN.
Thus the spectrum of electrons and positrons injected into ISM should be the
spectrum inside the PWN when it is disrupted, other than the spectrum inside the
magnetosphere of pulsar \cite{maly09}. Like SNRs, pulsars can be divided into local pulsars
and smooth distant components. \citet{dela10} find that the
contribution from pulsar background is negligible compared to those from local ones,
thus we do not take the former into account in our calculation. Parameters of
nearby pulsars can be found in the ATNF catalog \cite{atnf}.

We assume the injection $e^{\pm}$ spectra of PWNs have the same form of
Eq.~(\ref{eq:Q1D}); a burst-like injection spectrum is also adopted. Note that the
spectral index used here is associated with the PWN, and cannot be derived from the
spectral index of the pulsed radio emission from pulsar given in the ATNF
catalog. Thus if the radio spectral index of PWN is not available, we set
$\gamma_{\rm PWN}$ as a free parameter in the following fittings. The
normalization parameter $Q_0$ is linked to the spin-down energy $W_p$
dissipated by pulsar by:
\begin{equation}
\int_{E_{\rm min}}^{\infty}Q(E)E\,dE=\eta W_p\,,
\label{eq:pwn}
\end{equation}
where $E_{\rm min}=0.1\,$GeV and $\eta$ is the efficiency of energy conversion
treated as another free parameter. The spin-down luminosity
$\dot{E}$ evolves with the age of pulsar $t$ as
$\dot{E}=\dot{E_0}(1+t/\tau_0)^{-2}$ \cite{1973ApJ...186..249P}, where
$\dot{E_0}$ is the initial spin-down luminosity, $\tau_0$ is the spin-down time
scale of the pulsar, assumed to be 10~kyr in our work. Integrating $\dot{E}$
with time, we get the expression of spin-down energy
$W_p=\dot{E}\,t\,(1+t/\tau_0)$, where $\dot{E}$ and $t$ can be found in ATNF
catalog. For the cut-off energy,
We take $E_c=2\,$TeV following the work of Ref.~\cite{mauro14}.

\subsection{Secondary Electrons and Positrons}
Secondary electrons and positrons are created by inelastic collision between CR
nuclei and ISM. The CR nuclei mainly consist of protons and $\alpha$ particles
while H and He are major components of ISM. As can be seen from the AMS-02 result, the positron fraction is smaller than 10\% below
$\sim100\,$GeV, where secondary positrons should contribute less than 10\%.
Since
the spallation process produces more positrons than electrons, secondary
electrons possess a further smaller percentage comparing to the total electron
intensity in the range mentioned above. Thus we neglect the secondary electron
component and
concentrate on secondary positrons. In this part, our calculation follows the
method of Ref.~\cite{dela09}. The source function of secondary $e^+$ is assumed to
be steady and homogeneous in a slab geometry:
\begin{equation}
Q_{sec}(E) = 4\pi\sum_{i,j}n_j\int dE'\,\Phi_{i}(E')\,\frac{d\sigma_{ij}(E',E)}{dE}
\label{eq:qsec}
\end{equation}
so that the propagation equation can be solved semi-analytically with a
relatively
simple form (see \cite{dela09} for details). In Eq.~(\ref{eq:qsec}), $i$ and $j$ mark
the species of CR nuclei and ISM gas respectively. The number density of ISM is
set to be $n_{\rm H}=0.9\,\rm{cm^{-3}}$ and $n_{\rm He}=0.1\,\rm{cm^{-3}}$.
The intensities of incident CR nuclei are donated by $\Phi$ which can be
estimated
by the observed intensities at earth. We employ the expression of $\Phi$
proposed by \citet{shikaze07}. \citet{mauro14} fit the AMS-02 data of proton and Helium to refresh
the parameters in the model of $\Phi$. For differential scattering
cross-section $d\sigma/dE$, \citet{kamae06} provides functional formulae
for proton-proton collision. Empirical rescaling based on this result are
applied to estimate cross-section of collision between other species \cite{norbury07}.  Besides,
we introduce a free parameter $c_{e^+}$ to rescale the secondary
intensity, considering the uncertainty in the calculation above, to accommodate
the data.

\section{Fitting to the AMS-02 data}
\label{sec:ams}
Up to now, AMS-02 provides the most precise measurement of leptonic spectra
below 1~TeV. We attempt to perform global fittings to all the leptonic data
released by the AMS-02 Collaboration, including the positron fraction, positron plus
electron spectrum, positron spectrum and electron spectrum \cite{amsfrac,amselec,amstot}.
The global fitting is useful
to set stringent constraints for astrophysical contributors,
and leads to a self-consistent picture \cite{yuan15}.

Astrophysical components considered in our model have been discussed in the
previous section, including background SNRs as main contributors of electrons in low
energy range, local SNRs as dominant contributors of
electrons in higher energy range, secondary positrons which dominate low
energy part of positron spectrum, and local PWNs as
predominant components in higher energy range of positrons.

In this section, we explain the AMS-02 data by using several astrophysical
source models. The characteristic of each
model depends on which sources are chosen to be predominant local SNRs. We
can see below that according to the calculation in the previous section, local
SNRs listed in Table \ref{tab2} hardly have significant contributions to electron
intensity within the energy range of AMS-02. They would not play the role of
predominant local sources in this section; we simply add their contributions
for each
model.

Positrons contribute only $\sim\,10\%$ of the total
$e^-+e^+$ intensity. We would like to simplify the constitution of positron
spectrum, and use a single PWN to fit the high energy part of $e^+$ spectrum.
\citet{mauro14} have already given a 'single-source' analysis in their work and
find that Geminga is the most proper one among their candidates. However, due
to the old age of Geminga (342~kyr), a spectral roll-off may appear below
1~TeV. Thus if Geminga is chosen as the single PWN, it may induce a slight break
in $e^-+e^+$ spectrum below 1~TeV.

Another famous pulsar B0656+14, also namely Monogem, is believed to
be associated with Monogem Ring. It lies at a distance of 0.28~kpc with an age
of 112~kyr which is younger than Geminga \cite{atnf}. Monogem has the potential
to significantly contribute to the positron spectrum from 100~GeV to 1~TeV (see
Table 4 of Ref. \cite{dela10}). Also, we can see from Figure 2 of
Ref.~\cite{mauro14}, the spectrum of Monogem has a similar shape with that of
total PWNs in the ATNF catalog. Therefore we expect a PWN with the similar
distance and age as Monogem can play a role as the single positron source.

We would show in following subsections that Monogem does well in the fittings
as a single positron source.
Comparing with the total
spin-down energy of Geminga $1.26\times10^{49}\,$erg and the required $\eta$ of
only 0.27 in the 'single source' model in Ref. \cite{mauro14}, the fittings for Monogem requires
a large $\eta$ (0.6$\sim$0.8 in our models) due to its low spin-down energy of $1.78\times10^{48}\,$erg.
However, Monogem can extent its electron/positron spectra to higher energies than Geminga because of its smaller lifetime, and is helpful to explain the cut-off indicated by HESS and VERITAS data.
In this work, we take Monogem as the single positron source in our fittings.

Hence, for different models shown below, the common free parameters are $Q_0$
and $\gamma$ for SNR background, rescaling parameter $c_{e^+}$ for secondary
positron, $\eta$ and $\gamma_{\rm PWN}$ for Monogem and a solar modulation
potential $\phi$ to accommodate data below tens of GeV.

\begin{figure}
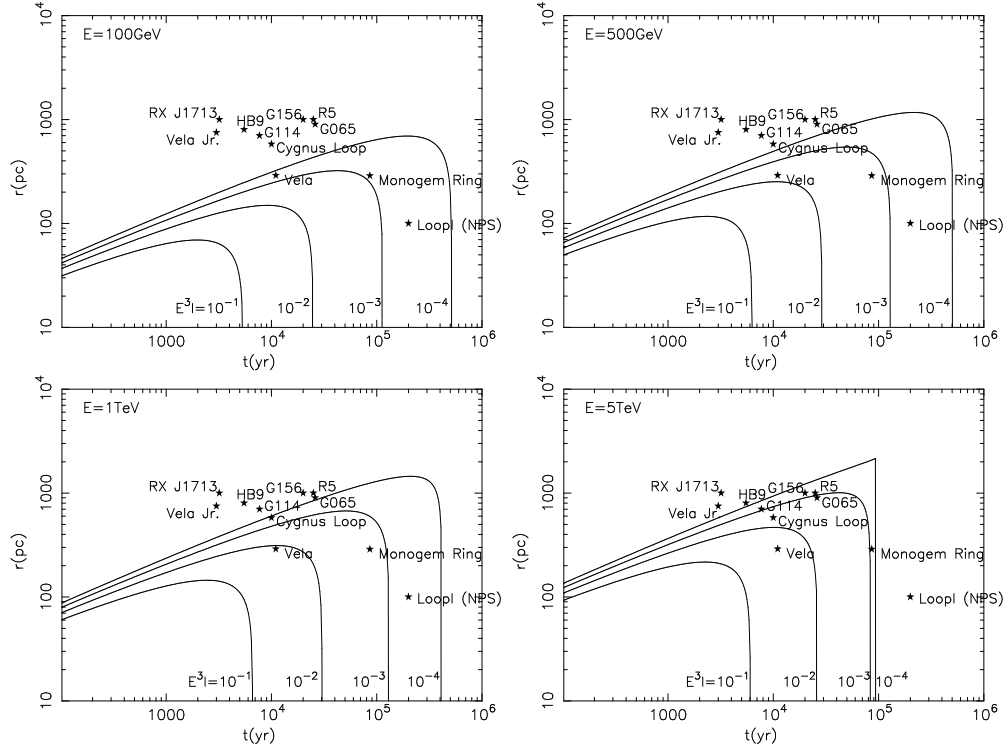

\centering
\includegraphics[width=0.3\textwidth, angle=270]{rt1.ps}
\includegraphics[width=0.3\textwidth, angle=270]{rt2.ps}
\includegraphics[width=0.3\textwidth, angle=270]{rt3.ps}
\includegraphics[width=0.3\textwidth, angle=270]{rt4.ps}
\caption{Contours of electron intensity as a function of ages and distances of
local SNRs. Top left: $E=100\,$GeV; top right: $E=500\,$GeV; bottom left:
$E=1\,$TeV; bottom right: $E=5\,$TeV. Contour lines in each graph represent for
$E^3I(E)=10^{-1},\,10^{-2},\,10^{-3},\,10^{-4}\,{\rm
GeV^2\,cm^{-2}\,s^{-1}\,sr^{-1}}$, as noted in each graph. In the calculation,
the spectral index, cut-off energy and total input energy of electrons are
assumed to be 2.0, 10~TeV and $10^{48}\,$erg, respectively. Sources listed in
Table \ref{tab1} are also marked in each graph. Since every SNR is constrained
to share
the same injection spectrum here, this figure is for reference only.}
\label{fig:rt}
\end{figure}

\subsection{Vela YZ}
It is widely believed that the famous and well studied SNR Vela is an
important local source of Galactic CRs due to its appropriate age and distance
and its strong radio emission \cite{koba04,mauro14}. Fig.~\ref{fig:rt} shows
the contours of electron
intensity as a function of $r$ and $t$ of source at different energy, assuming typical
input energy ($10^{48}\,$erg), spectral index (2.0) and cut-off energy
(10~TeV) of electrons. Every source in
Table \ref{tab1} is marked in Fig.~\ref{fig:rt}. It is clearly that Vela
predominates
over other local SNRs above hundreds of GeV, and this is where the
observed electron excess to continuous SNR model appears.

Vela shows a
shell-like radio structure consisting of three principle regions, dubbed
Vela X, Y and Z \cite{rish58}. A weaker component Vela W observed by
\citet{alvarez01} is not considered in our work. \citet{milne68} find the
spectrum of
Vela X was remarkably flat than that of Vela Y and Z. This unusual spectrum was
first explained by \citet{weiler80}. They argued that Vela X should belong to
plerions and PWNs like Crab, rather than shell-type SNRs like
its siblings Vela Y and Z. This point of view has been accepted in later
works. Consequently, when we estimate the contribution from Vela SNR, we ought
to divide Vela X from Vela YZ. PWNs have different mechanisms of acceleration
and evolution from shell-type SNRs. This means that they cannot share parameters,
such as spectral index or cut-off energy, with the later.

\begin{figure}
\centering
\includegraphics[width=0.65\textwidth, angle=270]{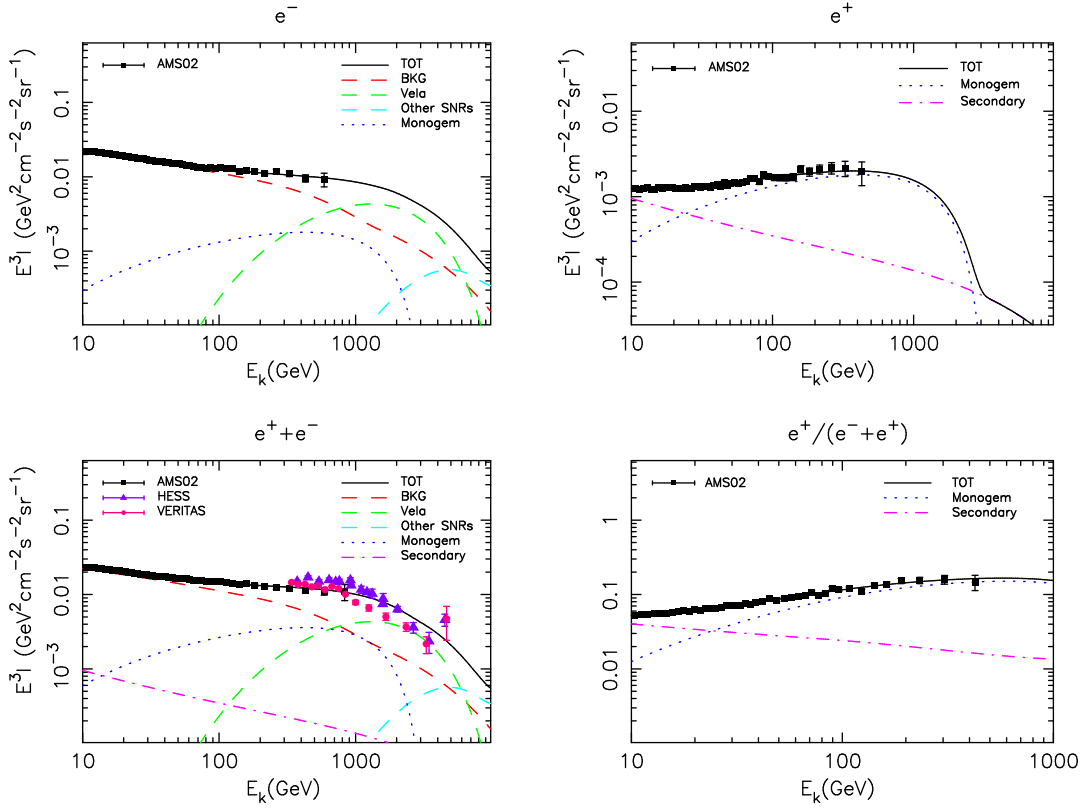}
\caption{The results of global fitting to AMS-02 data (\emph{Vela YZ} model).
Top left: electron
intensity;
top right: positron intensity; bottom left: positron plus electron intensity;
bottom
right: positron fraction. The legend explains different components in each
sub-graph. In the legends, 'TOT' stands for total value, 'BKG' stands for SNR
background, 'Other SNRs' refers to the summation of SNRs in Table II. Besides
AMS-02, data from HESS and VERITAS measurements are also shown in $e^++e^-$
graph.}
\label{fig:vela}
\end{figure}

In the first model, we set Vela YZ as the unique predominant SNR and aim to
check if it can give a good interpretation to AMS-02 data. As suggested by
Ref.~\cite{mauro14}, we fix the magnetic field to be 30~$\mu$G and cut-off
energy to be
2~TeV. In the Green catalog of SNRs, the radio spectral index of Vela is denoted
as 'varies' perhaps due to the discrepancy between Vela X and YZ. We put
spectral index $\alpha_{vela}$ of Vela YZ into the group of free parameters in
this model. A free $\alpha_{vela}$ leads to the uncertainty of $B_{r}^{\rm 1
GHz}$. The radio flux of Vela YZ at 960~MHz is measured to be 1100~Jy
\cite{alvarez01}, so we set $B_{r}^{\rm 1 GHz} = 1000$~Jy here as an estimation.
We
seek the best-fit model by minimizing chi-squared statics between model and data
points. The result of global fitting to AMS-02 data are shown in
Fig.~\ref{fig:vela}. In each sub-graph, all the components are drawn to show
their contribution and black solid line represents the fitting result. The
best-fit reduced $\chi^2$ is 0.473 for 182 degrees of freedom (d.o.f.), while
the best-fit parameters are compiled in Table \ref{tab3}. Solar modulation $\phi$ converges to zero in this case.

As can be seen that this model well explains the AMS-02 data. The
best-fit radio spectral index of Vela YZ seems reasonable, as it is close to the
typical value 0.5. \citet{mauro14} also fit the four leptonic observables of
AMS-02
simultaneously and obtain fairly good result. However, our model has two main
differences from theirs. First, the propagation parameters they adopted is the
\emph{MED} model proposed by Ref.~\cite{donato04} which is based on the B/C
analysis
performed in Ref.~\cite{maurin01}, and the nuclei data behind is taken from
earlier
balloon and space experiments; as described in Sec.~\ref{subsec:ppg}, we
include the latest B/C data from AMS-02 in our calculation of propagation
parameters. Besides, Vela X
and YZ are assumed to provide a whole SNR contribution in their fittings. As mentioned above, these objects
may not share the same spectral index, cut-off energy, and normalization
parameter $Q_0$. If the contribution of Vela X is considered in the energy range covered by AMS-02, it
should have contributed to positron intensity due to its PWN nature, and would not induce a spectral structure at high energies above TeV.

\begin{sidewaystable}
\raggedright
\begin{tabular}{|c|c|c|c|c|c|c|c|c|}
  \hline
  Model & $\gamma$ & $Q_0[10^{50}{\rm GeV^{-1}}]$ & $c_{e^+}$ & $\gamma_{\rm PWN}$ & $\eta$ & $\phi [\rm GV]$ & $\alpha_{vela}$ & $\chi^2/{\rm d.o.f}$\\
  \hline
  Vela & $2.411^{+0.005}_{-0.006}$ & $1.77^{+0.04}_{-0.03}$ & $0.745^{+0.033}_{-0.034}$ & $1.98^{+0.08}_{-0.07}$ & $0.742^{+0.044}_{-0.043}$ & 0 & $0.519^{+0.32}_{-0.06}$ & 0.473 \\
  \hline
\end{tabular}
\begin{tabular}{|c|c|c|c|c|c|c|c|c|c|c|}
  \hline
  Model & $\gamma$ & $Q_0[10^{50}{\rm GeV^{-1}}]$ & $c_{e^+}$ & $\gamma_{\rm PWN}$ & $\eta$ & $\phi [\rm GV]$ & $\alpha_{mr}$ & $W_{mr}$[$10^{48}\,$erg] & $E_{c,mr}$[TeV] & $\chi^2/{\rm d.o.f}$ \\
  \hline
  ${\rm Vela+MR}_{\alpha=0.53}$ & $2.602^{+0.006}_{-0.006}$ & $3.52^{+0.08}_{-0.08}$ & $0.956^{+0.042}_{-0.042}$ & $1.89^{+0.09}_{-0.08}$ & $0.646^{+0.049}_{-0.048}$ & $0.351^{+0.099}_{-0.095}$ & $0.551^{+0.061}_{-0.063}$ & $2.89^{+0.45}_{-0.44}$ & $1.04^{+14.73}_{-0.74}$ & 0.398 \\
  \hline
  ${\rm Vela+MR}_{\alpha=0.735}$ & $2.569^{+0.007}_{-0.005}$ & $3.17^{+0.06}_{-0.08}$ & $0.930^{+0.042}_{-0.038}$ & $1.90^{+0.09}_{-0.07}$ & $0.656^{+0.048}_{-0.047}$ & $0.304^{+0.104}_{-0.084}$ & $0.484^{+0.096}_{-0.082}$ & $2.13^{+0.40}_{-0.40}$ & $1.13^{+11.01}_{-0.78}$ & 0.396 \\
  \hline
\end{tabular}
\begin{tabular}{|c|c|c|c|c|c|c|c|c|c|c|}
  \hline
  Model & $\gamma$ & $Q_0[10^{50}{\rm GeV^{-1}}]$ & $c_{e^+}$ & $\gamma_{\rm PWN}$ & $\eta$ & $\phi [\rm GV]$ & $\alpha_{loop}$ & $W_{loop}$[$10^{48}\,$erg] & $E_{c,loop}$[TeV] & $\chi^2/{\rm d.o.f}$ \\
  \hline
  ${\rm Vela+Lp1}_{\alpha=0.53}$ & $2.579^{+0.007}_{-0.007}$ &
$2.95^{+0.07}_{-0.07}$ & $0.898^{+0.040}_{-0.040}$ & $1.92^{+0.09}_{-0.08}$ &
$0.669^{+0.048}_{-0.048}$ & $0.256^{+0.106}_{-0.100}$ &
$0.438^{+0.028}_{-0.106}$ & $4.94^{+0.58}_{-0.57}$ & $1.02^{+0.97}_{-0.47}$ &
0.401 \\
  \hline
  ${\rm Vela+Lp1}_{\alpha=0.735}$ & $2.580^{+0.007}_{-0.007}$ &
$3.00^{+0.07}_{-0.07}$ & $0.907^{+0.040}_{-0.040}$ & $1.91^{+0.09}_{-0.08}$ &
$0.665^{+0.048}_{-0.048}$ & $0.265^{+0.104}_{-0.098}$ &
$0.417^{+0.068}_{-0.060}$ & $5.08^{+0.62}_{-0.62}$ & $1.49^{+1.32}_{-0.65}$ &
0.400 \\
  \hline
\end{tabular}
\caption{Fitting results of models presented in Sec. \ref{sec:ams}. Loop I
is abbreviated to Lp1 in this table.}
\label{tab3}
\end{sidewaystable}

\subsection{Vela YZ + Monogem Ring}
In this model, we attempt to investigate the parameters of Vela YZ at first, rather
than taking typical values as in the previous model. The geometry of Vela SNR
can be sketched by two hemisphere with different scales due to the asymmetrical
density distribution of its surrounding medium \cite{sushch14}. Vela Y and Z
are located in the
north-east(NE) part. Assuming the equilibrium between thermal pressure and
magnetic pressure, \citet{sushch14} give an estimation of $B_{NE}=46\,\mu G$
with
the formula $B_{NE}=\sqrt{8\pi n_{NE}k_BT_{NE}}$, where density $n_{NE}$ and
temperature $T_{NE}$ are derived by Ref.~\cite{sushch11}.  We still cite
Eq.~(\ref{eq:Ec}) to calculate cut-off energy. But instead of taking typical
value $10^8\,{\rm cm\,s^{-1}}$, we choose $v_{NE}=6\times10^7\,{\rm cm\,s^{-1}}$
estimated by Ref.~\cite{sushch11}. The principle behind this estimation is the
expression
of shock radius as a function of the age of SNR given by Ref.~\cite{white91}.
We get
$E_c=4\,$TeV. For $\alpha_r$ of Vela YZ, \citet{dwa91} claims a value of
0.53 combining his 34.5~MHz observation and other observations up to 2700~MHz.
The latest work about radio spectrum of Vela is done in Ref.~\cite{alvarez01}.
The authors of Ref.~\cite{alvarez01} included more data points at other frequencies while subtracted the flux
density at 34.5~MHz for the probable absorption of this measurement. The
$\alpha_r$ of Vela YZ is derived to be 0.735 in this work. For $\alpha_{vela}=0.53$,
$B_{r}^{\rm 1GHz}$ is derived to be 800~Jy; while for $\alpha_{vela}=0.735$, we
derive $B_{r}^{\rm 1GHz}\sim700\,$Jy.

\begin{figure}
\centering
\includegraphics[width=0.65\textwidth, angle=270]{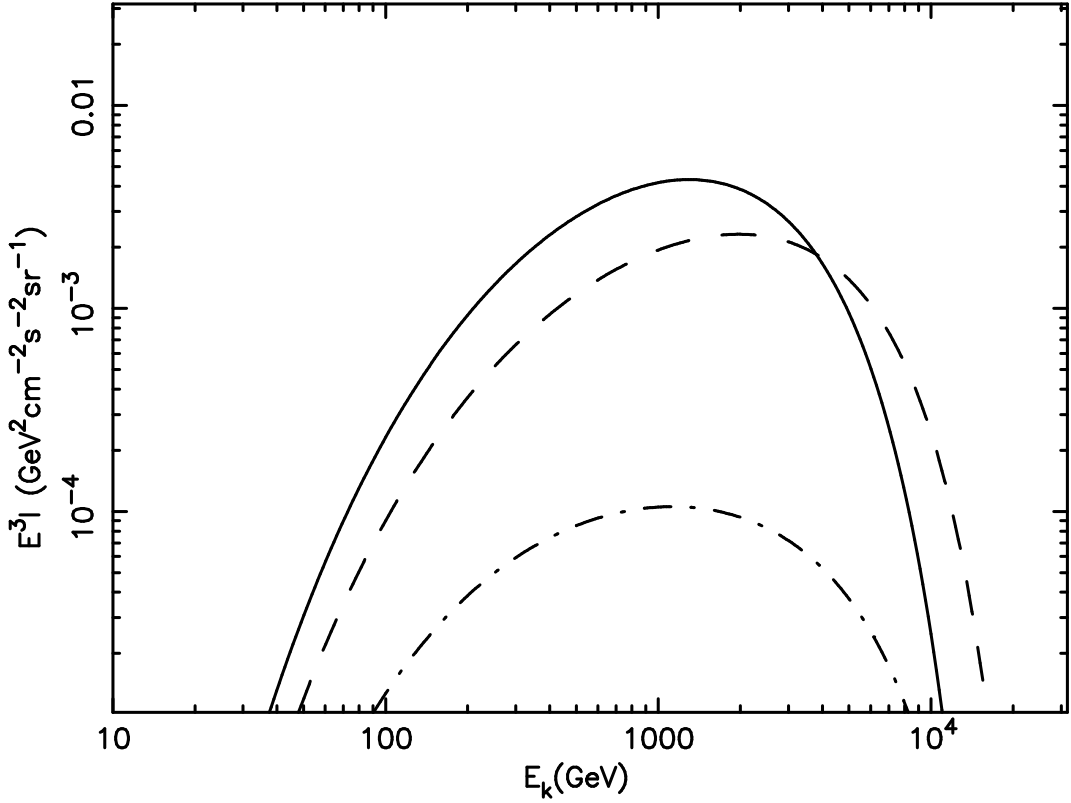}
\caption{Electron spectra of Vela YZ with different parameters. The solid line
describes the case $B=30\,\mu$G, $E_c=2\,$TeV and $\alpha_{vela}=0.519$. The
first two are typical values and the last is given by fitting. The dashed line
and the dot-dashed line come from $B=46\,\mu$G, $E_c=4\,$TeV,
$\alpha_{vela}=0.53$ for the former and $\alpha_{vela}=0.735$ for the later.}
\label{fig:3vela}
\end{figure}

Fig.~\ref{fig:3vela} shows the comparison between the electron spectra of Vela
YZ with parameters in the previous model (solid line) and parameters in this
model (dashed line for $\alpha_{vela}=0.53$ and dot-dashed line for
$\alpha_{vela}=0.735$). If we take our parameters estimated above and still use
Vela YZ as the only predominant SNR in the fitting, excesses of $e^-$ and
$e^-+e^+$ data would arise in hundreds of GeV, especially for the case
$\alpha_{vela}=0.735$. Thus, we need to find a proper SNR as a cooperator in
this case. Fig.~\ref{fig:vela} indicates SNRs listed in Table \ref{tab2}
hardly have
influence on electron intensity in the energy range of AMS-02, as we mentioned
above. From Fig.~\ref{fig:rt}, we can see Monogem ring and
Loop I are the potential electron contributors up to 1~TeV. We would like to
examine Monogem ring in this model.

Monogem Ring (MR) is a large
shell-like structure in soft X-ray band with a diameter $\sim25^\circ$
\cite{plucinsky96}. Green
does not include large X-ray regions with scales larger than $10^\circ$ in his
catalog \cite{green}, thus MR does not appear there. Assuming MR is in its
adiabatic phase,
\citet{plucinsky96} derives parameters of MR with observable quantities and its
distance, applying Sedov-Taylor model of SNR. \citet{plucinsky09} points out
300~pc
should be a reasonable approximation for distance of MR, and consequently
an age of $8.6\times10^4\,$yr and an initial explosion energy
$0.19\times10^{51}\,$erg. Apart from this, we possess poor knowledge of MR to
constrain its radio spectral index, total energy converted into electrons, or
cut-off energy. We add these three quantities, $\alpha_{mr}$, $W_{mr}$ and
$E_{c,mr}$ to free parameters in this scenario.

\begin{figure}
\centering
\includegraphics[width=0.65\textwidth, angle=270]{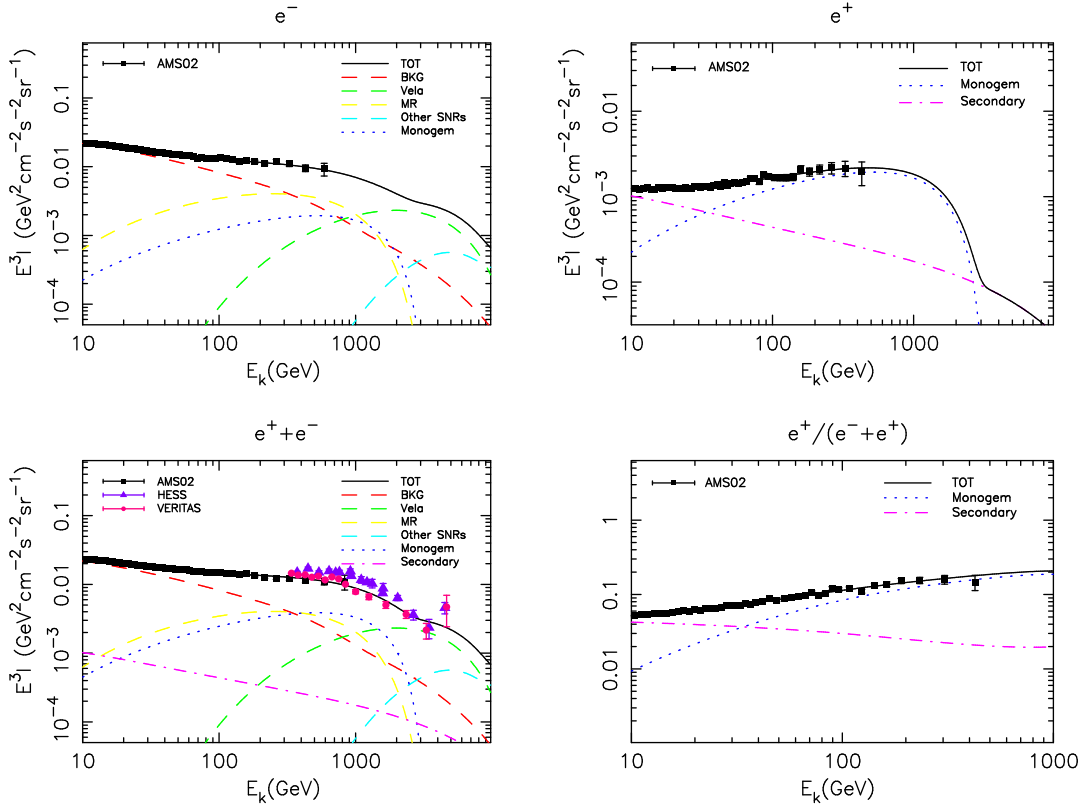}
\caption{Same as Fig.~\ref{fig:vela}, but for \emph{Vela + Monogem Ring}
scenario
with $\alpha_{vela}=0.53$. In the legends, 'MR' is the abbreviation for Monogem
Ring.}
\label{fig:mr1}
\end{figure}

\begin{figure}
\centering
\includegraphics[width=0.65\textwidth, angle=270]{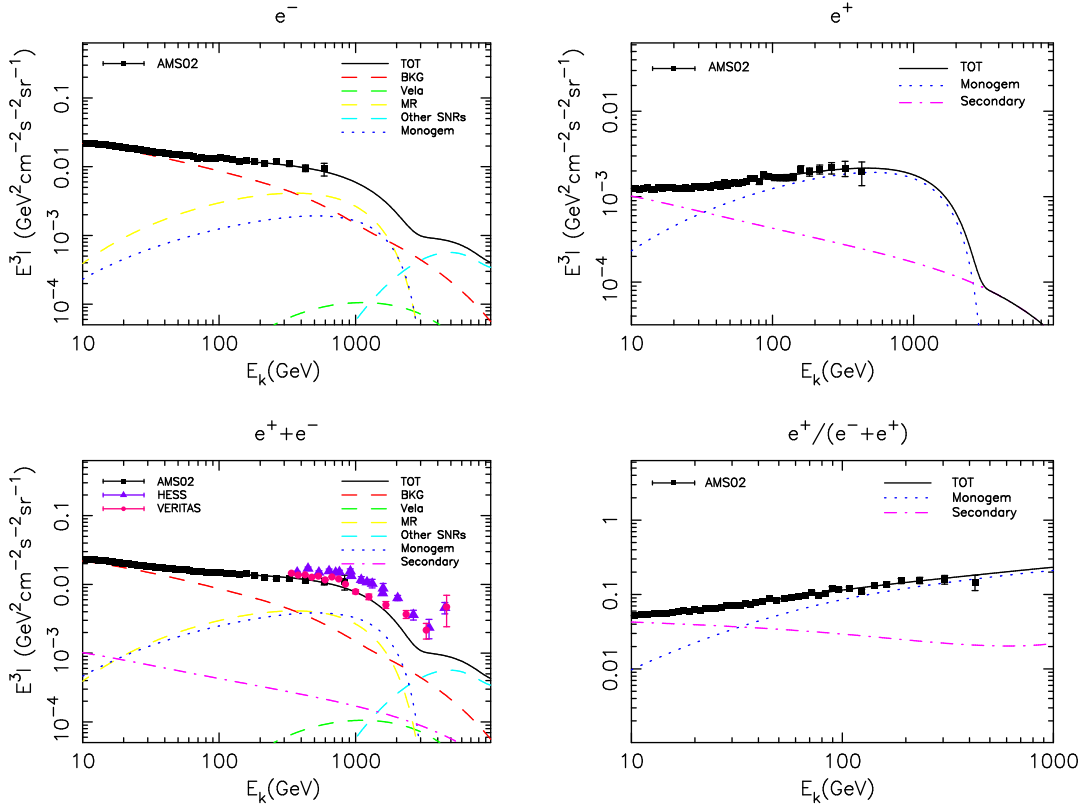}
\caption{Same as Fig.~\ref{fig:vela}, but for \emph{Vela + Monogem} Ring
scenario
with $\alpha_{vela}=0.735$. In the legends, 'MR' is the abbreviation for
Monogem
Ring.}
\label{fig:mr2}
\end{figure}

Two global fittings are done with $\alpha_{vela}=0.53$ and
$\alpha_{vela}=0.735$, and fitting results are plotted in Fig.~{\ref{fig:mr1}}
and Fig.~{\ref{fig:mr2}} respectively. For the case $\alpha_{vela}=0.53$, the
best-fit reduced $\chi^2$ is 0.398 for 180 d.o.f; if we take the latest value
$\alpha_{vela}=0.735$, the result becomes $\chi^2/{\rm d.o.f}=0.396$ for 180
d.o.f. Best-fit parameters are shown in Table \ref{tab3} for both cases. It can
be seen, good fitting
results to AMS-02 data are achieved when MR joins in the model. In $e^-$ and
$e^{-}+e^{+}$ spectra, the deviation between these two results with different
$\alpha_{vela}$ becomes clear in TeV range.

\subsection{Vela YZ + Loop I (NPS)}
In this model, we largely repeat the work of the previous model, but replace
MR by Loop I. It is well known that there are several Galactic giant
radio loops potentially associating with SN events \cite{berk71,
berk73,duncan97}. Loop I, which was discovered
by \citet{large62} and named by \citet{berk71b}, is the most prominent one
among
these loops. Although Loop I, with a radius of $58^{\circ}$, is also not
included in the Green's catalog, it has gone through more careful study than
MR. The center of Loop I is close to Scorpio-Centaurus OB association
where SN events happen. North Polar Spur (NPS) is the most prominent part of
Loop I, both in radio and X-ray maps \cite{egger95}. However, the visible X-ray
contradicts
with the age of $10^6\,$yr derived by the low expanding velocity of H I
surrounding Loop I \cite{bunner72,sofue74}. Then a
reconciliation is raised: Loop I is indeed
an old structure, but it has been reheated by younger SN events
\cite{borken77,egger95}. The shock wave
from the most recent SN event in Sco-Cen association happened in $2\times10^5$
years ago and gave rise to the X-ray feature of NPS \cite{egger95}. The
comparison between
the continuous shell-like radio structure and the interrupted X-ray structure
of Loop I also favors this interpretation. Thus $2\times10^5\,$yr should be
take as the age of Loop I (NPS) in our work, and adopting the
corresponding distance to NPS---100~pc \cite{bingham67}---as the distance of
this source is more
reasonable than taking that of Sco-Cen association (170~pc) as in some earlier
researches.  Similar to the previous model, free parameters associating with
Loop I (NPS) in the following fittings are $\alpha_{loop}$, $W_{loop}$ and
$E_{c,loop}$.

\begin{figure}
\centering
\includegraphics[width=0.65\textwidth, angle=270]{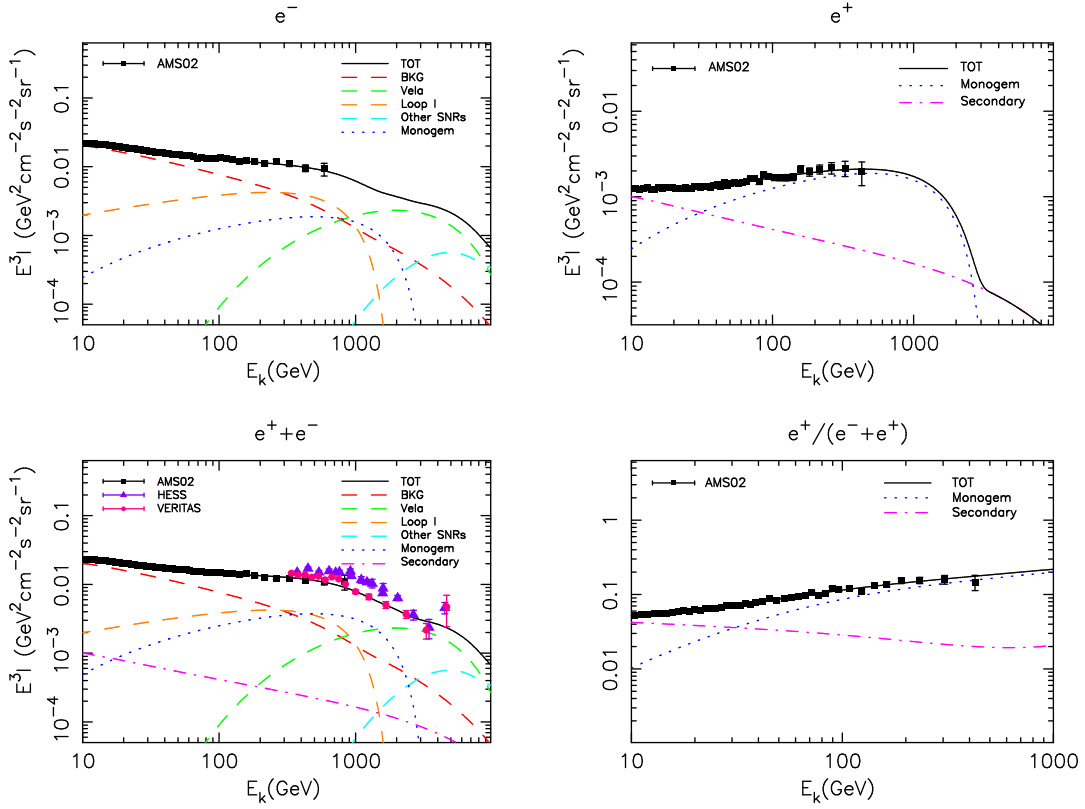}
\caption{Same as Fig.~\ref{fig:vela}, but for \emph{Vela + Loop I (NPS)}
scenario
with $\alpha_{vela}=0.53$. }
\label{fig:loop1}
\end{figure}

\begin{figure}
\centering
\includegraphics[width=0.65\textwidth, angle=270]{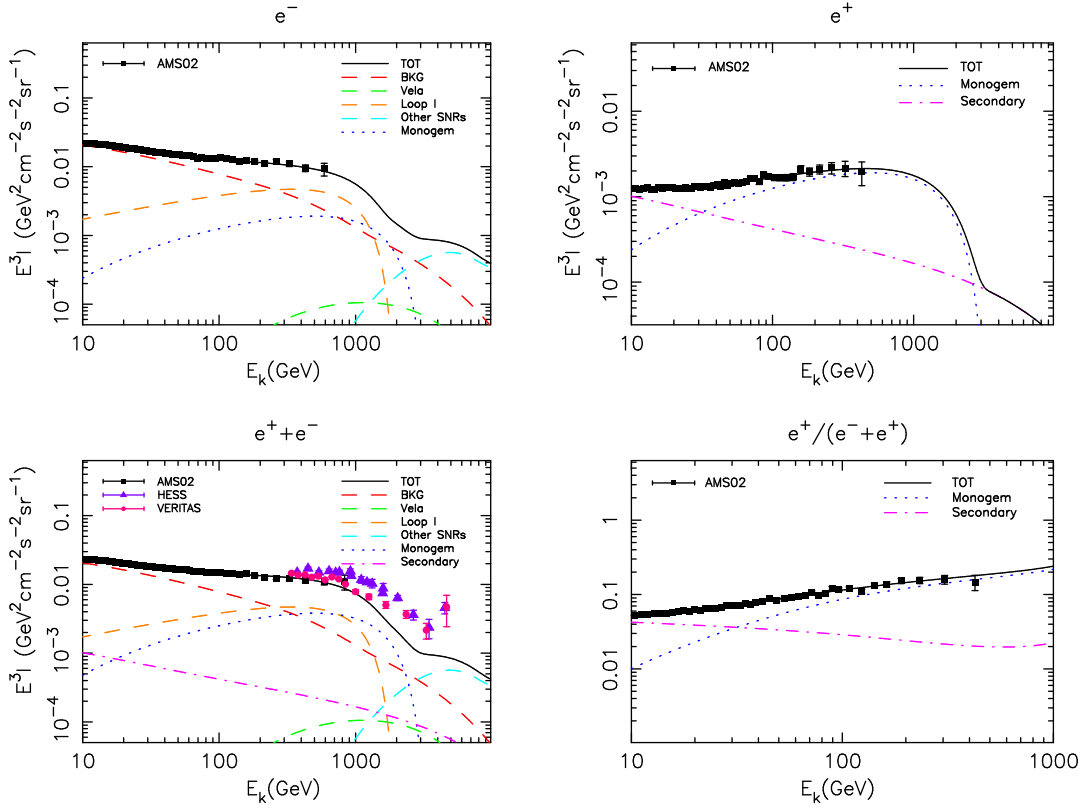}
\caption{Same as Fig.~\ref{fig:vela}, but for \emph{Vela + Loop I (NPS)}
scenario
with $\alpha_{vela}=0.735$. }
\label{fig:loop2}
\end{figure}

The fitting results are shown in Fig.~\ref{fig:loop1} and Fig.~\ref{fig:loop2}
for the case of $\alpha_{vela}=0.53$ and $\alpha_{vela}=0.735$, respectively.
The best-fit reduced $\chi^2$ for the former case is 0.401 for 180 d.o.f,
while for the later case, $\chi^2/{\rm d.o.f}=0.400$ for 180 d.o.f. See Table
\ref{tab3} for corresponding best-fit parameters. Comparing with the reduced
$\chi^2$ of the previous model, the fitting effect of this \emph{Vela YZ + Loop
I} scenario has little difference with that of \emph{Vela YZ + Monogem Ring}
scenario.

\subsection{Discussion}
In this section, we have proposed several source models to fit the leptonic
spectra measured by AMS-02. It can be seen that the radio spectral index of Vela YZ is
crucial to the situation. Radio measurements from Wilkinson Microwave
Anisotropy Probe (WMAP) and Planck may give further constraints
on $\alpha_{vela}$,
but processed data for Vela is not available at present. If the radio spectral
index of Vela YZ is $\sim$0.5, other local
SNRs, may not be necessary to join in fitting to the AMS-02 data, as in the \emph{Vela
YZ} model. However, if we
choose a value of 0.735, Vela loses its predominance of the
local contribution. Since the
electron intensity contributed by the PWN is confined by $e^+$ spectrum, other local
SNRs, such as MR and Loop I, are needed to reproduce the observed $e^-$ spectrum.

In the \emph{Vela YZ + Monogem Ring} model, the best fitting lepton energy is
$2.13\times10^{48}\,$erg, near the typical
value of $10^{48}\,$erg. However,
the low explosion energy of MR derived by \citet{plucinsky09} favors a small
input energy
of electrons, or a very large conversion efficiency of $\sim10^{-2}$ is
required. For the \emph{Vela YZ + Loop I} model, we get a larger input energy
for electrons of $5.08\times10^{48}\,$erg. However, this large value seems not
excluded by the initial SN energy of Loop I of $10^{52}\,$erg,
which may be
produced by several SN events \cite{egger95}. Our fitting result of
$\alpha_{loop}$ is 0.417,
which is smaller than the value of 0.5 given by the radio observation to NPS between
22 and 408 MHz \cite{roger99} or between 45 and 408 MHz \cite{guzman11}.
The measurements between 408 and 1420
MHz get a even larger value of 1.1 \cite{reich88}. It is also possible that both
MR and Loop I
play important roles in the $e^-$ spectrum. This setting may relax the ranges of $W_{mr}$ and
$\alpha_{loop}$ somewhat.

\section{Electron spectrum above TeV}
\label{sec:tev}

Now we turn attention to higher energy part beyond the scope of AMS-02.
DAMPE is expected to detect electrons/positrons in the range of 5~GeV to 10~TeV
\cite{dampe}.
Models proposed in the previous section have already
given $e^-+e^+$ spectra extending to TeV range, which may be measured
in future DAMPE data. However, although these models give different
predictions in TeV range depending on the characteristic of Vela YZ, a common
decreasing is shown in $e^-+e^+$ spectrum up to 10~TeV in each
model. This means that these models predict no protruding structure in the highest energy
range of DAMPE.

Ground-based Cherenkov telescopes, such as HESS \cite{hess08,hess09}, MAGIC \cite{magic} and
VERITAS \cite{veritas}, have extended the measurement of $e^-+e^+$ spectrum to several TeV.
These measurements show that a spectral steepening appears above 1~TeV.
Furthermore, HESS and the preliminary result of VERITAS have measured a
coincident ascending of $e^-+e^+$ intensity in $\sim5\,$TeV, which may imply a
feature from local sources. Unfortunately, this tendency comes from only the
most energetic data point in both case and no measurement has been taken above
5~TeV. Thus it is the show time for DAMPE to examine this tendency.

In this
section, we give additional predictions to electron spectrum above TeV. We would
study whether local sources can produce distinctive features in the
high energy range covered by DAMPE. Although we do not intend to fit the data of
HESS, MAGIC, or VERITAS quantitively, we need to keep in mind that the spectral
steepening just beyond 1~TeV could be reproduced in our models.

\subsection{Vela X}
There is still an important source which have not been included in our models
so far: Vela X, as the sibling of Vela YZ. Vela X is one of the most well
studied PWN
powered by PSR B0833--45 \cite{weiler80}. It has been covered by observations from radio bands
to very high energy $\gamma$-ray bands \cite{alvarez01,markwardt97,mangano05,abramowski12,abdo10,grondin13}.  In morphology, Vela X consists of an
extended halo and a small collimated fearture embeded in the halo, e.g., the
'cocoon'. TeV $\gamma$-ray emssion has been detected in the cocoon region by
HESS, while Fermi-LAT observations has reported the presence of sub-GeV-peak
$\gamma$-ray emssion extending in the halo. \citet{jager08} proposes a model
with two populations of electrons to interpret this phenomenon: a high energy
component concentrating on the cocoon responsible for X-ray and TeV
$\gamma$-ray emssion, and a lower energy component extending in the halo
responsible for radio and GeV $\gamma$-ray emssion. Following this idea,
assuming the leptonic origin of $\gamma$-ray emssion, \citet{abdo10} give a
multi-wavelength fitting to the SED of Vela-X. In their results, the cut-off
energy of the halo is only 100~GeV which is too low to help Vela-X appearing in
TeV range; for the cocoon, although its cut-off energy is high enough, a total
lepton energy of $1.5\times10^{46}\,$erg is too weak to produce significant
structure in TeV spectrum.

\begin{figure}
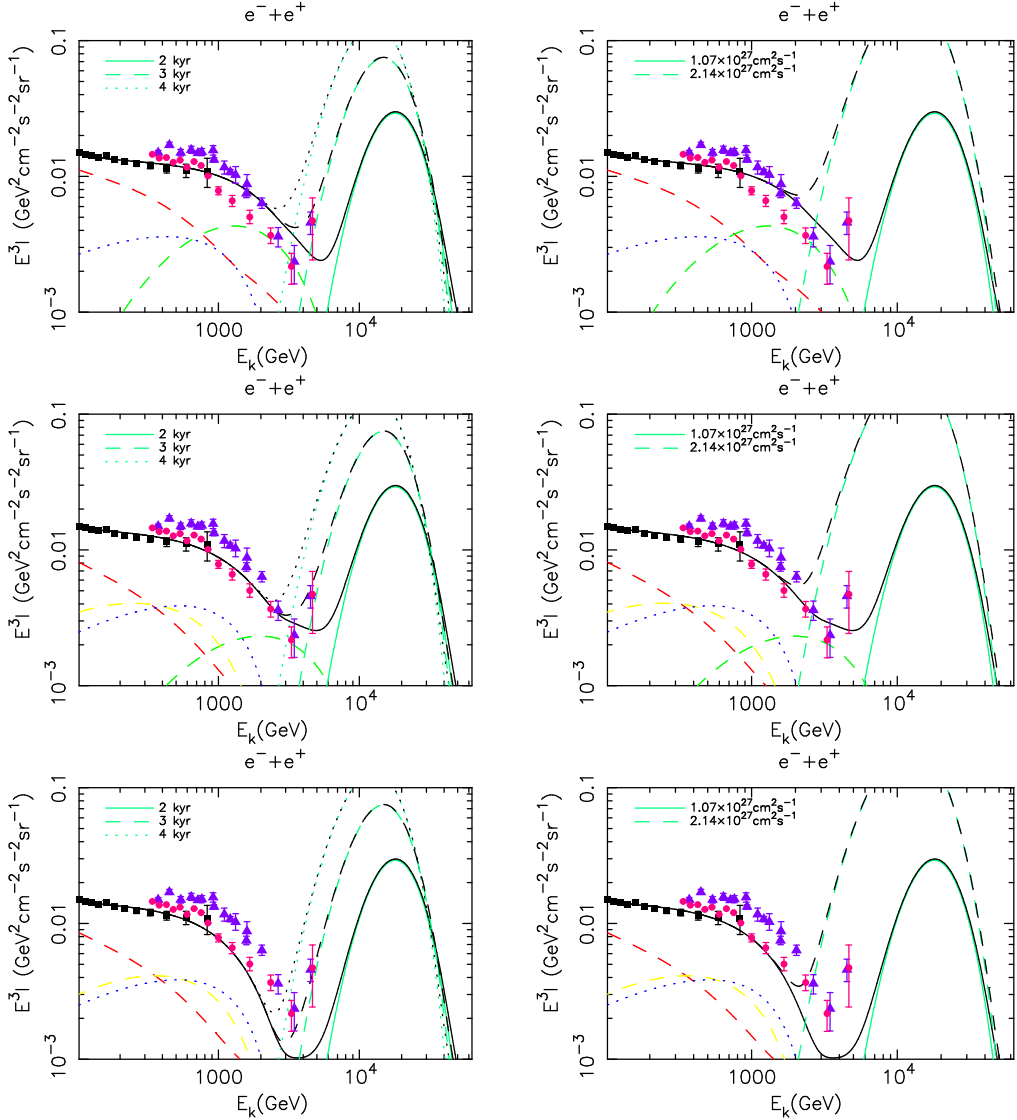

\centering
\includegraphics[width=0.3\textwidth, angle=270]{velax1.ps}
\includegraphics[width=0.3\textwidth, angle=270]{velax2.ps}
\includegraphics[width=0.3\textwidth, angle=270]{velax3.ps}
\caption{Predictions to $e^-+e^+$ spectrum above TeV combining models in
Sec.~\ref{sec:ams} and Vela X as predominant contributor in TeV range. Top
left and top right: Vela X combining with \emph{Vela YZ} model; middle left and
middle right: Vela X combining with \emph{Vela YZ + Monogem Ring} model
($\alpha_{vela}=0.53$); bottom left and bottom right: Vela X combining with
\emph{Vela + Monogem Ring} model ($\alpha_{vela}=0.735$). The legends in
Fig.~\ref{fig:vela}, \ref{fig:mr1} and \ref{fig:mr2} are still valid in this
figure to describe corresponding components. Left panels show effects of varying
the injection age of Vela X and right panels show effects of changing $D_0$. }
\label{fig:velax}
\end{figure}

However, as mentioned above, these electron features derived by photon
emission may not describe those electrons we have observed. \citet{hinton11}
provide an alternative model in which a serious particle
escape has happened in the halo while particles in the cocoon are well
confined. After staying in confinement status for a long time,
PWN begins to interact with the coming reverse shock of SNR. It is the time
that PWN crushes and burst-like injection happens. From then on, the halo has
been suffering from energy-dependent escape, thus the cut-off energy derived by
Fermi-LAT observations is such a small value. From some moment after the PWN
crushes, the pulsar starts to inject particles to a new PWN, that is, the
cocoon. This explains the dimness of the cocoon for its short time of particle
injection.

\citet{hinton11} gives an estimation of electron injection spectrum:
$E^{1.8}{\rm exp}(E/6\,{\rm TeV})$ with an total energy $6.8\times10^{48}\,$erg.
Thus Vela X seems to produce a TeV feature. However, the diffusion
coefficient adopted by them ($1.07\times10^{27}(E/1\,{\rm GeV})^{0.6}\,{\rm
cm^{2}\,s^{-1}}$ for energy much larger than 1\,GeV) is almost an order of
magnitude smaller than ours. If we take our $D(E)$, the intensity from Vela X
is much larger, especially at lower energy. The spectral steepening
indicated by HESS and VERITAS cannot be reproduced, and even the energy range
of AMS-02 would be affected. Since the diffusion scale is proportional to
$\sqrt{D(E)t}$, a larger $D(E)$ means a faster propagation and more electrons
with lower energy are able to arrive at the Earth. If we want to keep the
steepening feature in 1~TeV, an unreasonable young age of Vela X is needed.

We combine Vela X and each model in the previous section to give new
predictions. As discussed above, we have to keep the $D(E)$ given by
Ref.~\cite{hinton11} only for Vela X. Fig.~\ref{fig:velax} shows the predicted
$e^-+e^+$ spectra. Since there is little difference between the spectral
shape of \emph{Vela YZ + MR} model and \emph{Vela YZ + Loop I} model, we only draw the
former case as representative. The effects of varying injection time of Vela X
and diffusion coefficient are also shown in Fig.~\ref{fig:velax}. We take three
different injection age of Vela X as 2~kyr, 3~kyr, and 4~kyr. Note these ages are
still observed ages, keeping consistent with those listed in Table \ref{tab1}. With
injection age increasing, the steepening structure in 1~TeV is gradually
filled up. This requires an injection age less than 5~kyr, according with the
theory that reverse shock has recently crushed the PWN \cite{blondin01}. Moreover,
Fig.~\ref{fig:velax} indicates the spectrum of Vela X depends sensitively on
diffusion coefficient (we double the $D_0$ for comparison). With the data of
DAMPE and a clearer picture of the particle escape of PWN in the future, Vela X
may become a potential tool to constrain diffusion coefficient in high energy.

\begin{figure}
\centering
\includegraphics[width=0.65\textwidth, angle=270]{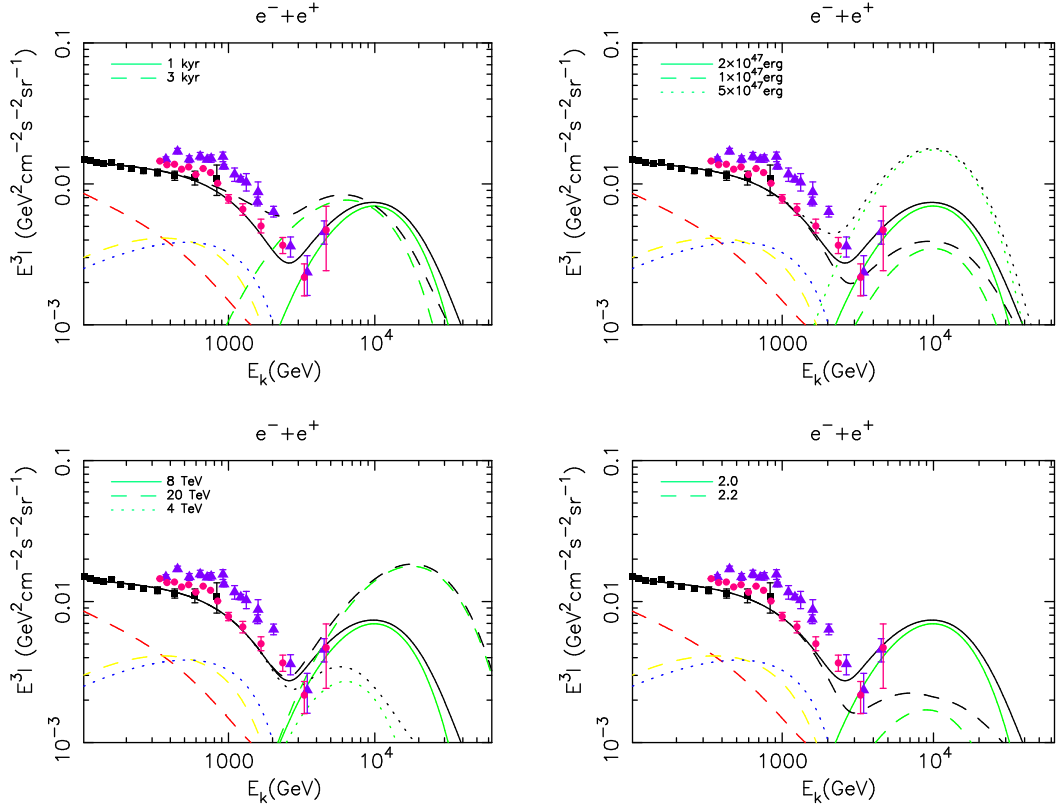}
\caption{Predictions to $e^-+e^+$ spectrum above TeV replacing the Vela YZ
in \emph{Vela YZ + Monogem Ring} model ($\alpha_{vela}=0.735$) with the Vela YZ
estimated in Sec.~\ref{subsec:velayz}. Top left: effect of varying observed
age of Vela YZ; top right: effect of varying input energy of leptons of Vela
YZ; bottom left: effect of varying cut-off energy of Vela YZ; bottom right:
effect of varying electron spectral index of injection spectrum of Vela YZ.
Black lines in each graph stand for total intensity corresponding to the Vela
YZ with same line style, and solid green line in each graph represents Vela
YZ adopting our fiducial parameters:  $t=1\,$kyr, $W=2\times10^{47}\,$erg,
$E_c=8\,$TeV and $\gamma=2.0$. The legends in Fig.~\ref{fig:mr2} are still
valid in this figure to describe corresponding components.}
\label{fig:velayz}
\end{figure}

\subsection{Vela YZ}
\label{subsec:velayz}
Naturally, following Vela X, the next question is if Vela YZ can play an
important role in the highest energy part of DAMPE. The work of
Ref.~\cite{koba04} shows the possibility of this scenario after considering the
release time of electrons in SNR, although they do not make a distinction
between Vela X and Vela YZ. \citet{erlykin02} believe that electrons start their
propagation after the expansion phase of SNR which hold a typical time scale of
200~yr. This time delay is too small to change the electron spectrum of
Vela YZ. Alternatively, \citet{dorfi00} point out particles begin to escape from
the shock front when the SNR dissolves in the ISM, i.e., when the velocity of
the shock has dropped to the mean Alfv$\rm{\acute{e}}$n velocity of ISM.
However, even we take a very low number density of $0.01\,{\rm cm^{-3}}$, the
derived release time is several times of $10^4\,$yr which is larger than
the observed age of Vela YZ or the typical time scale of Sedov phase. To insure
electrons have already escaped from Vela YZ, we should control its release time
not longer than $10^4\,$yr, corresponding to a least observed age 1~kyr. The
total input energy of Vela YZ is much smaller than that of Vela X, it should be
at the magnitude of $10^{47}\,$erg. When the electron acceleration is synchrotron
loss limited, the cut-off energy decreases with the age growth of SNR. Since
$E_c=4\,$TeV is derived by observations of electromagnetic radiation, the
injection spectrum of electrons should have a larger cut-off energy. Similarly,
higher energy electrons should suffer heavier energy loss, which causes
softening of spectrum with the growth of trapped time of electrons. Thus the
spectral index of injection spectrum may smaller than that derived by radio
observations.

We set $t=1\,$kyr, $W=2\times10^{47}\,$erg, $E_c=8\,$TeV and
$\gamma=2.0$ as fiducial values and use some other values of these parameters to
show spectral variation in Fig.~\ref{fig:velayz}. In this figure, we apply
\emph{Vela YZ + Monogem Ring} model ($\alpha_{vela}=0.735$) in
Sec.~\ref{sec:ams}.
In this model Vela YZ can bring ascending or relatively flat spectral features
just below 10~TeV, as shown in Fig.~\ref{fig:velayz}. We can imply from this
figure that if the radio spectral index is taken 0.735 here, Vela YZ will have
no chance to produce spectral feature in $\sim\,$TeV.

\subsection{Vela Jr.}
\begin{figure}
\centering
\includegraphics[width=0.65\textwidth, angle=270]{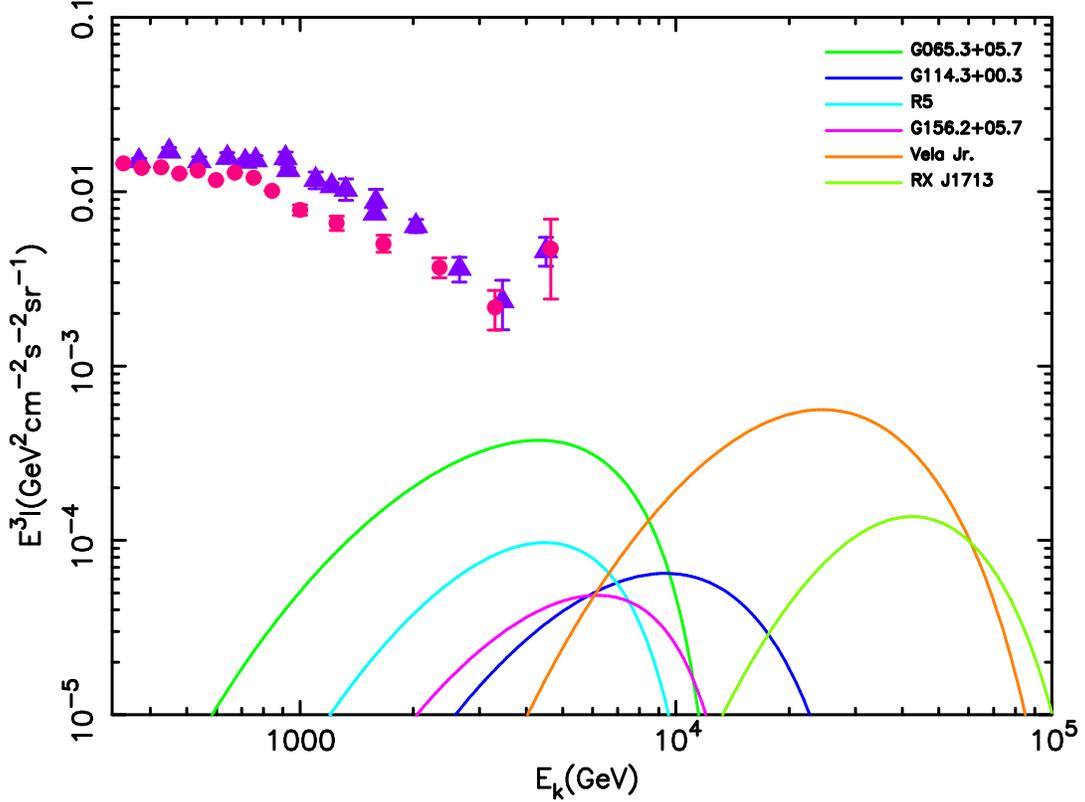}
\caption{Electron intensities of individual local SNRs listed in Table
\ref{tab2}.
Cygnus loop and HB9 do not appear in the scope of this figure because of their
very low cut-off energy, as shown in Table \ref{tab2}. Data from HESS and
VERITAS
measurements are also shown in this figure.}
\label{fig:local}
\end{figure}

Besides Vela XYZ, other candidates producing prominent structure in the TeV range
should be selected from Table \ref{tab2}. Fig.~\ref{fig:local} shows intensity of each SNR listed in Table II, along with the data of HESS and VERITAS. Cygnus Loop and HB9 do not appear in the scope of Fig.~\ref{fig:local} because of their
very low cut-off energy. To affect the $e^-+e^+$ spectrum in $\sim\,$TeV, Vela
Jr. seems to need least parametric adjustment, namely, a larger input energy to leptons. Thus we examine Vela Jr. as an example.

Vela Jr. locates
in the southeastern corner of Vela on the sky map, but at a farther distance of
750~pc. Vela Jr. is one of those sources whose parameters are estimated
by fitting the multi-wavelength emission in Sec.~\ref{subsubsec:snr_para}.
\citet{tanaka11} and \citet{lee13} also conducted broadband analysis to Vela Jr. in recent years. As there have been discussions about magnetic field of Vela Jr.,
$Q_0$ and $E_c$ can be estimated by Eq.~(\ref{eq:Q0}) and Eq.~(\ref{eq:Ec}).
We put aside our fitting result to parameters of Vela Jr. temporarily.
Chandra has detected spindly filamentary structure in Vela Jr. \cite{bamba05}. This thin
structure is interpreted by efficient synchrotron cooling of CR electrons in a
strong local magnetic field of $\sim\,100\,\mu$G \cite{berezhko09}. However, if broadband
emission is modeled by leptonic scenario, a magnetic field of $\sim\,10\,\mu$G
is required to explain the synchrotron to inverse Compton flux ratio. Thus we
take three different magnetic fields of $100\,\mu$G, $10\,\mu$G and a typical
value $30\,\mu$G adopted in Ref.~\cite{mauro14}. Here $\alpha_{r}$ and $B_r^{1{\rm
GHz}}$ are taken to be 0.3 and 50~Jy, as given in Table \ref{tab1}, to calculate $Q_0$. We take the upper limit of the age of Vela Jr., i.e. an observed age of
4.3~kyr.

\begin{figure}
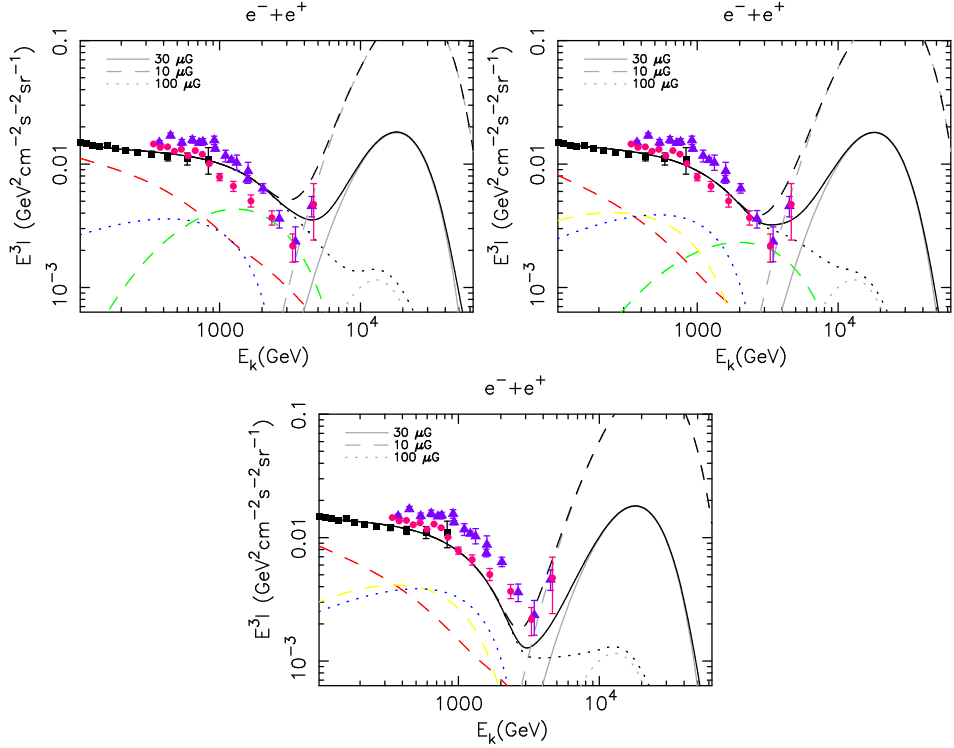

\centering
\includegraphics[width=0.3\textwidth, angle=270]{velajr1.ps}
\includegraphics[width=0.3\textwidth, angle=270]{velajr2.ps}
\includegraphics[width=0.3\textwidth, angle=270]{velajr3.ps}
\caption{Predictions to $e^-+e^+$ spectrum above TeV combining models in
Sec.~\ref{sec:ams} and Vela Jr. as predominant contributor in TeV range.
Top left: Vela Jr. combining with \emph{Vela YZ} model; top right: Vela Jr.
combining with \emph{Vela YZ + Monogem Ring} model
($\alpha_{vela}=0.53$); bottom: Vela Jr. combining with
\emph{Vela + Monogem Ring} model ($\alpha_{vela}=0.735$). The legends in
Fig.~\ref{fig:vela}, \ref{fig:mr1} and \ref{fig:mr2} are still valid in this
figure to describe corresponding components. Effects of adopting different
magnetic field are shown in these graphs. Black lines in each graph stand for
the total intensity corresponding to the Vela Jr. with the same line style.}
\label{fig:velajr}
\end{figure}

The predicted $e^-+e^+$ spectrum of Vela Jr. combining with \emph{Vela YZ} and
\emph{Vela YZ + Monogem Ring} models (of course original Vela Jr. subtracted)
are shown in Fig.~\ref{fig:velajr}. The
input leptonic energy corresponding to magnetic field of $30\,\mu$G,
$100\,\mu$G and $10\,\mu$G are $1.72\times10^{48}\,$erg,
$2.80\times10^{47}\,$erg and $8.92\times10^{48}\,$erg, respectively. For the
case of $B=100\,\mu$G, Vela Jr. cannot produce prominent feature in
$\sim\,$TeV; when $B=10\,\mu$G, the spectral feature given by Vela Jr. is
remarkable enough, but it needs a total leptonic energy much higher than the
typical value of $10^{48}\,$erg. The result given by $B=30\,\mu$G may be what
we'd like to see. However, we should note the $\alpha_r$ of Vela Jr. is derived
by the radio flux of only two wave bands, which is quite unreliable, and the
value of $\alpha_r$ will give a significant influence to the total leptonic
energy of Vela Jr. with the method used here. In fact, the parameters given by
the leptonic model of Ref.~\cite{tanaka11} is similar to the result of our broadband
fitting (the differences may be due to the consideration of the energy loss
between injection electrons and radiation electrons in the work of
\cite{tanaka11}). \citet{lee13} also found the leptonic model is clearly superior to
hadronic model. Thus broadband fitting may give better constraint to parameters
of Vela Jr., which disfavor Vela Jr. as a prominent contributor in $\sim\,$TeV.

\begin{figure}
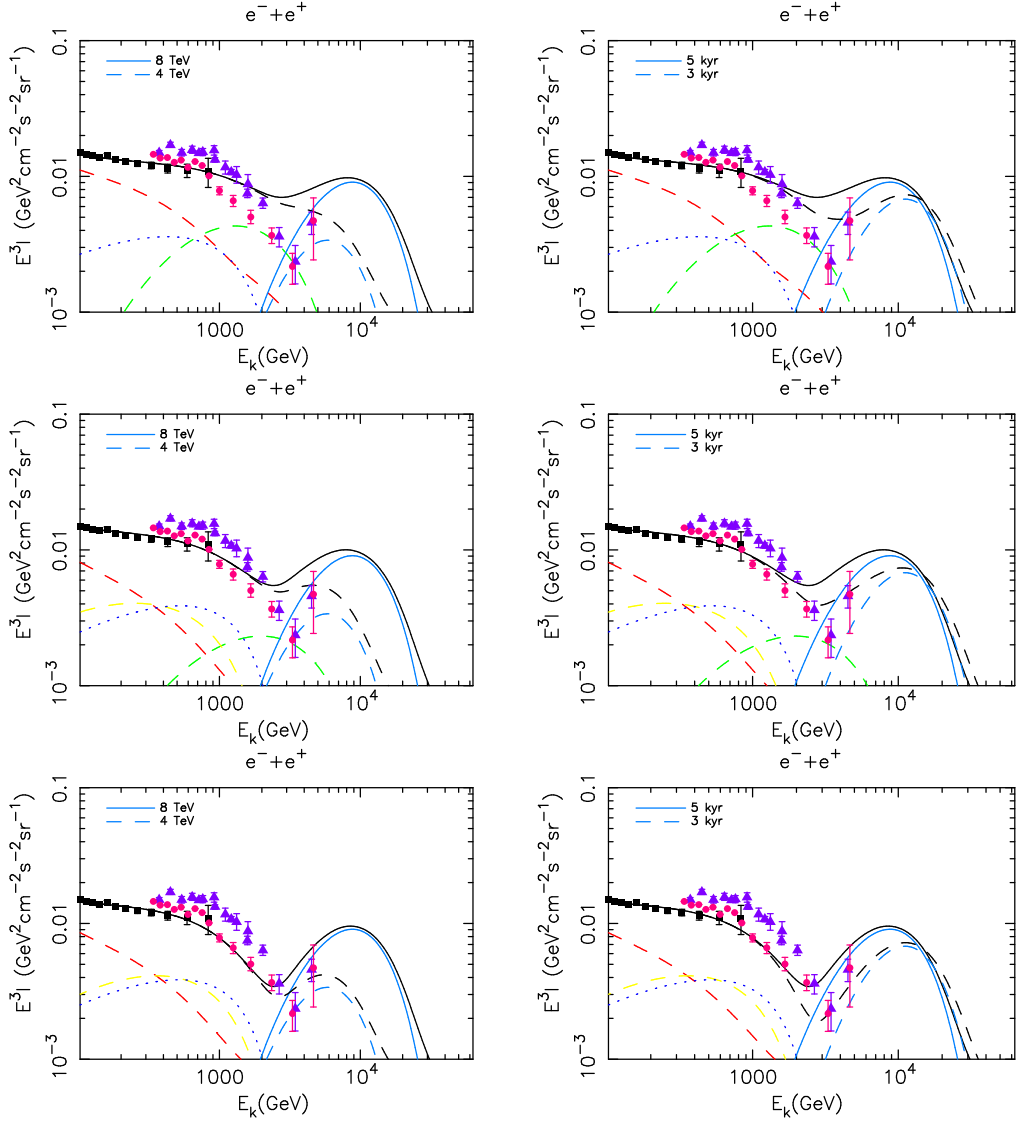

\centering
\includegraphics[width=0.3\textwidth, angle=270]{cygnus1.ps}
\includegraphics[width=0.3\textwidth, angle=270]{cygnus2.ps}
\includegraphics[width=0.3\textwidth, angle=270]{cygnus3.ps}
\caption{Same as Fig.~\ref{fig:velax}, but replace Vela X with Cygnus Loop. Left
panels show effects of varying the cut-off energy of Cygnus Loop and right
panels show effects of changing the injection time. }
\label{fig:cygnus}
\end{figure}

\subsection{Cygnus Loop}
Cygnus Loop is one of the closest SNRs to the Earth (540~pc) and is considered
as an important local CR accelerator second only to Vela \cite{mauro14}. However, its steep GeV
spectrum observed by Fermi-LAT (see Fig.~\ref{fig:broadband}) indicates a very low cut-off energy
of electron spectrum, thus it cannot play a part in the models above. Cut-off
energy derived by Eq.~(\ref{eq:Ec}) is at least several TeV for Cygnus Loop, so
it is possible that Cygnus Loop has undergone serious particle escape similar to
the case of Vela X. Unfortunately, although Cygnus Loop has been carefully
studied in X-ray band with the help of XMM-Newton and Suzaku \cite{uchida08,leahy13}, its global X-ray spectrum
is not available. Thus we cannot give further inference from the broadband
spectrum of Cygnus Loop as Vela X. Here we assume cut-off energy of TeV for
Cygnus Loop, which is the precondition for Cygnus Loop to contribute
significantly in TeV range. Besides, to preserve the spectral steepening in
1~TeV, considerable release time is needed for Cygnus Loop. We keep others
parameters given by broadband fitting unchanged. Like the former
cases, we combine Cygnus Loop with models given in Sec.~\ref{sec:ams} and
also show the effects of varying cut-off energy and observed age of Cygnus Loop
in Fig.~\ref{fig:cygnus}. In the left panels, observed age is fixed at 5~kyr,
while in the right panels, cut-off energy of Cygnus Loop is set to be 8~TeV.

\subsection{Discussion}
As discussed in the last section, it is hard to tell which local SNR
dominates the electron
excess in the energy range of AMS-02, since the spectra of local SNRs are mixed
with that of the SNR background below 1~TeV.
However, things are different above the
TeV scale that can be covered by DAMPE.
The contributions from the background SNRs may be very small in this energy range. If we get a distinctive spectrum above several TeV, we may
determine the origin of these cosmic ray electrons.
Of course, it is also possible that more than one
sources shape the high energy electron spectrum, then the situation becomes
complicated.

In order to produce clear features at the spectrum beyond TeV,
a small particle propagation time and a high energy cut-off of injection electrons
are essential. We have considered the Vela X, Vela YZ, Vela Jr. and
Cygnus Loop as possible candidates contributing to cosmic electrons above TeV in
this section. Except for very young sources like Vela Jr., other three
candidates need an additional release time. We find that Vela X and Vela Jr. may
provide a very sharp rise
in the spectrum at a few TeV due to the very young components,
while the Vela YZ and Cygnus Loop may show much smoother feature at the similar
energy. Thus with the basis of a $\sim$ kyr injection age and a $\sim$ TeV cut-off
energy, the total leptonic injection energy of a source is the key factor to produce a sharp
spectral structure in the high energy range covered by DAMPE.

\section{Summary}
In this paper, we have given predictions to CR electron (plus positron)
spectrum above $\sim$ TeV basing on the elaborate analysis of the local astrophysical sources,
especially the local SNRs. In order to obtain a complete picture, we ensure the
consistency between the predicted spectra and present experimental results below TeV by performing global fittings to all the latest leptonic AMS-02
data. We find that Vela YZ could act as the dominator just below TeV because of its proper
age and distance. However, it should be emphasized that the determination of the injection spectral
index of Vela YZ, which still has some uncertainties, is crucial to fittings. We discuss different scenarios to
fit AMS-02 data corresponding to different values of the spectral index of Vela YZ. Other SNRs, such as Monogem Ring and Loop I, are also introduced, if Vela YZ dose not provide the dominant contribution to the AMS-02 electron results.

Basing on the fitting results, we discuss the parameters of several local
sources, and give further expectations of the electron spectrum above TeV.
Adopting different
possible values for those parameters, we predict either sharp or relatively
flat electron spectral features, comparing with the monotonic decreasing spectra of the models in Sec. \ref{sec:ams}. All these models are ready for the examination at high
energy CR electron detectors, such as DAMPE, which can reach the energy as high as $\sim10$
TeV. The spectrum measurement
of DAMPE and future anisotropy measurements \cite{2016arXiv161106237M} may reveal
the origin of the high energy CR electrons. These results
will also be important for probing the mechanism of
CR acceleration in the sources.

\acknowledgments{This work is supported by the National Natural Science Foundation of China
under Grants No.~11475189,~11475191, and by the 973 Program of China under
Grant No.~2013CB837000, and
by the National Key Program for Research and Development (No.~2016YFA0400200).
}

\bibliography{electron}

\end{document}